\documentclass[a4paper,10pt,prl,twocolumn,superscriptaddress]{revtex4-1}
\usepackage{color,amsmath,amssymb,amsthm,times,graphics,graphicx,bm,bbm,dcolumn}

\newcommand{\be}{\begin{equation}}
\newcommand{\ee}{\end{equation}}

\def\tr{ {\rm{Tr }}}

\begin{document}
\title{Coherent open-loop optimal control of light-harvesting dynamics}

\author{Filippo Caruso}
\affiliation{Institut f\"{u}r Theoretische Physik, Universit\"{a}t
  Ulm, Albert-Einstein-Allee 11, D-89069 Ulm, Germany}

\author{Simone Montangero}
\affiliation{Institut f\"{u}r Quanteninformationsverarbeitung,
  Universit\"{a}t Ulm, Albert-Einstein-Allee 11, D-89069 Ulm, Germany}

\author{Tommaso Calarco}
\affiliation{Institut f\"{u}r Quanteninformationsverarbeitung,
  Universit\"{a}t Ulm, Albert-Einstein-Allee 11, D-89069 Ulm, Germany}

\author{Susana F. Huelga}
\affiliation{Institut f\"{u}r Theoretische Physik, Universit\"{a}t
  Ulm, Albert-Einstein-Allee 11, D-89069 Ulm, Germany}

\author{Martin B. Plenio}
\affiliation{Institut f\"{u}r Theoretische Physik, Universit\"{a}t
  Ulm, Albert-Einstein-Allee 11, D-89069 Ulm, Germany}
\begin{abstract}
We apply theoretically open-loop quantum optimal control techniques
to provide methods for the verification of various quantum coherent
transport mechanisms in natural and artificial light-harvesting
complexes under realistic experimental constraints. We demonstrate
that optimally shaped laser pulses allow to faithfully prepare the
photosystem in specified initial states (such as localized excitation
or coherent superposition, i.e. propagating and non-propagating states)
and to probe efficiently the dynamics. These results provide a path towards the
discrimination of the different transport pathways and to the
characterization of environmental properties, enhancing our understanding of the role that coherent processes may play in biological complexes.
\end{abstract}
\maketitle
\section{Introduction}
Recent experimental work has provided evidence supporting the existence of
long-lived electronic coherence during excitation energy transfer (EET) in
photosynthetic complexes \cite{fleming07a,Engel10}. Subsequent theoretical
work has then highlighted the importance of an intricate interplay of noise
and quantum coherence for the efficiency of excitation energy transfer (EET)
in light harvesting complexes during photosynthesis and identified the crucial
building blocks that underly this interplay
~\cite{Aspuru08,PlenioHuelga08,ccdhp09,castro08,ccdhp10,patrick,cdchp10,chp10}.
While computer simulations and analytical work allow us to identify and verify
the importance of these effects in theory, the experimental verification of
their relevance in actual bio-molecular systems is still outstanding.
One path forward towards this goal is the application of optimal control methods
to the bio-molecular quantum dynamics in contact with environments with the aim
of preparing specific initial states and control the subsequent dynamics. This
would then allow for the generation of dynamical behaviour and hence signals that
can lead to the biggest discrepancies between alternative theoretical hypotheses.
Indeed, here we employ optimal control to develop different strategies that,
when experimentally tested, will allow to enhance our comprehension of
coherent processes in biological complexes.

Quantum coherent control drives a quantum system dynamics towards a specific goal
by exploiting quantum coherence and interference effects~\cite{rabitz}. Coherent
control techniques have been proposed for photochemical and
photobiological processes - see Refs.~\cite{shapiro,coh-review,pierce,kosloff,rice,may} for an overview on this topic. For instance, shaped light has been used to
discriminate spectroscopically indistinguishable biochromophores through
selective fluorescence depletion~\cite{petersen} and also to control energy flow
in bacterial photosynthesis by increasing the amount of a triplet state signal
in comparison to a singlet state signal in carotenoids~\cite{buckup}. The first
evidence of control of exciton states in light-harvesting systems was presented
experimentally in Ref.~\cite{herek}. Moreover, to the best of our knowledge, the
first attempt to use open-loop optimization to control the exciton dynamics of
a light-harversting system was theoretically provided in Ref.~\cite{brueggemann},
however the control algorithm employed there imposed some limitations to their
analysis: in particular their optimization was limited to state populations.

To overcome these limitations, we present and apply a recently introduced
optimization algorithm (CRAB)~\cite{DCM}.
In  particular, we apply quantum (open loop) optimal quantum control theory to
the dynamics of the electronic excitations in the Fenna-Matthews-Olson (FMO), a
biological pigment-protein complex involved in the early steps of photosynthesis
in green sulphur bacteria~\cite{fleming07a,Engel10,fleming07b,Prokhorenko,olson}.
Specifically, by using the CRAB approach in the context of the FMO complex
i) we achieve general state preparation: this will allow us to prepare specific
initial states, especially fast and slow propagating states exhibiting constructive
or destructive interference, ii) we explore experimental constraints and
imperfections (adapted to the expected experimental setup \cite{exp}), iii) we optimize
the difference in signals for different preparations to test theoretical
hypotheses, and iv) we also discuss optimized probing of the system by
optimal control of the readout pulses. Finally we provide an outlook discussing
various possible extensions of our approach.

\section{The Model}
\begin{figure}[t]
\centerline{\includegraphics[width=.5\textwidth]{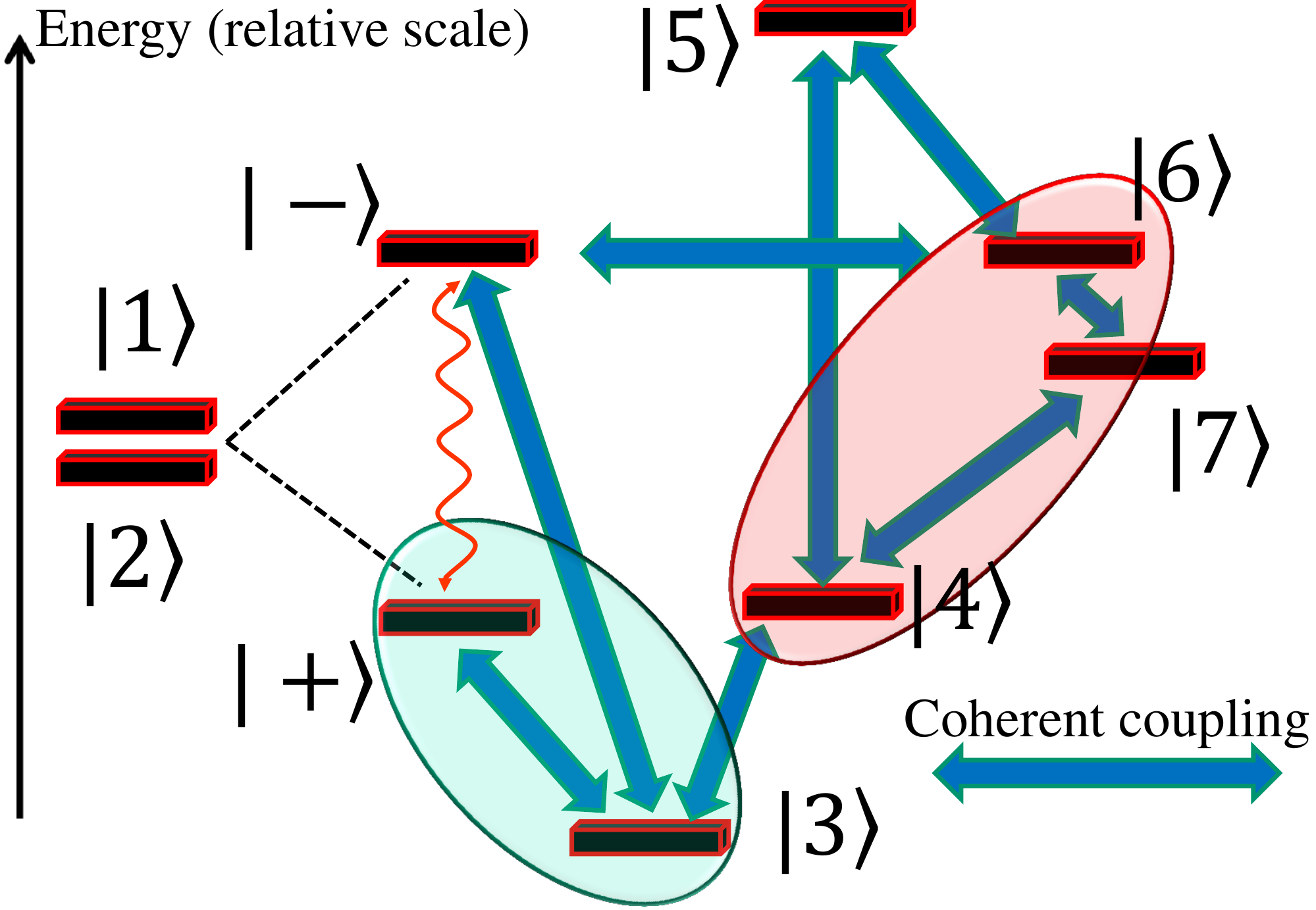}}
\caption{FMO energy level structure where $|i\rangle$ denotes a single excitation in site $i$. The states $|\pm\rangle$ are
the symmetric and anti-symmetric superpositions of the states $|1\rangle$ and $|2\rangle$. The green and red bubbles identify the fast and slow transport path, as detailed in Ref.~\cite{cdchp10} where the interplay between different transport pathways in FMO dynamics was discussed at length. The main effect of the inclusion of dephasing noise is the opening of an incoherent relaxation channel from level $|-\rangle$  to level $|+\rangle$  (red wiggled line) and therefore the effective suppression of the coherent oscillation between level $|-\rangle$ and sites
$6$-$7$-$4$ that dominates the coherent dynamics and is responsible of the very slow transport once the sink population has reached $50\%$ (initial population held in site $|+\rangle$). The proposed quantum control strategies efficiently probe this dynamical model.}
\label{fig1}
\end{figure}

In this section we present the basic ingredient of our theoretical
model and fix the notation. The effective dynamics of the FMO complex
can be modeled by a $7$-qubit Hamiltonian describing the coherent
exchange of excitations between chromophores or sites, i.e.
$$H_{fmo} = \sum_{j=1}^7 \hbar\omega_j \sigma_j^{+}\sigma_j^{-}
        + \sum_{j\neq l} \hbar v_{j,l} (\sigma_j^{-}
        \sigma_{l}^{+} + \sigma_j^{+}\sigma_{l}^{-})$$
where $\sigma_j^{+}$ ($\sigma_j^{-}$) are the raising (lowering) operators
for site $j$, $\hbar\omega_j$ is the local site excitation energy, and
$v_{j,l}$ denotes the hopping rate of an excitation between the sites
$j$ and $l$ - see Ref. ~\cite{ccdhp09} for more details about this model.
In the site basis, we follow \cite{adolphs06} and employ the Hamiltonian
matrix elements
\begin{equation}
        H \!=\!\! \left(\!\!\begin{array}{rrrrrrr}
         215   & \!-104.1 & 5.1  & -4.3  &   4.7 & -15.1 &  -7.8 \\
        \!-104.1 &  220.0 & 32.6 & 7.1   &   5.4 &   8.3 &   0.8 \\
           5.1 &   32.6 &  0.0 & -46.8 &   1.0 &  -8.1 &   5.1 \\
          -4.3 &    7.1 &\!-46.8 & 125.0 &\! -70.7 &\! -14.7 &  -61.5\\
           4.7 &    5.4 &  1.0 & \!-70.7 & 450.0 &  89.7 &  -2.5 \\
         -15.1 &    8.3 & -8.1 & -14.7 &  89.7 & 330.0 &  32.7 \\
          -7.8 &    0.8 &  5.1 & -61.5 &  -2.5 &  32.7 & 280.0
          \end{array}\!\!
        \right)
        \label{hami} \nonumber
\end{equation}
where the zero of energy has been shifted by $12230$ $\text{cm}^{-1}$ for all
sites, corresponding to a wavelength of $\cong 800~\mathrm{nm}$ (all
numbers are given in units of $\text{cm}^{-1}=1.988865\cdot 10^{-23}~\mathrm{Nm}
= 1.2414 \ 10^{-4}~\mathrm{eV}$) -- see Fig. \ref{fig1}.
In a first approximation, and with the aim of identifying the main transport
paths, the dissipation and dephasing caused by the surrounding environment are
modeled by the following local Lindblad terms
\begin{eqnarray}
    {\cal L}_{diss}(\rho) &=& \sum_{j=1}^{7} \Gamma_j
    [-\{\sigma_j^{+}\sigma_j^{-},\rho\} + 2 \sigma_j^{-}\rho \sigma_j^{+} ] \\
{\cal L}_{deph}(\rho) &=& \sum_{j=1}^{7} \gamma_j[-\{\sigma_j^{+}\sigma_j^{-},\rho\} + 2 \sigma_j^{+}\sigma_j^{-}\rho \sigma_j^{+}\sigma_j^{-}] \; ,
        \end{eqnarray}
with $\Gamma_j$ and $\gamma_j$ being the dissipative and dephasing rates at
the site $j$, respectively.
In the following, we will consider the case in which the dephasing and dissipation
rates are equal for all sites and labeled. As in Ref. ~\cite{ccdhp09}, we denote
the common rates as $\gamma=\gamma_j$ and $\Gamma=\Gamma_j = 5 \times
10^{-4}~\mathrm{ps}^{-1}$ for site $j=1,\ldots,7$. The latter corresponds to the
measured lifetime of excitons which is of the order of $1~\mathrm{ns}$. Finally,
the transfer efficiency into the reaction center is measured in terms of the
population in the `sink', numbered $8$, which is populated by an irreversible
decay process (with rate $\Gamma_{sink}$) from the site $3$, as described by
the Lindblad term
\begin{eqnarray}
        {\cal L}_{sink}(\rho) &=& \Gamma_{sink}[2\sigma_{8}^{+}\sigma_3^{-}
        \rho \sigma_3^{+}\sigma_{8}^{-} - \{\sigma_3^{+}\sigma_{8}^{-}\sigma_{8}^{+} \sigma_3^{-},\rho\} ] \; ,
        \label{sink}
\end{eqnarray}
where $\Gamma_{sink} \sim 6.3 ~ \mathrm{ps}^{-1}$ (note that $\hbar \sim 5.3$ cm$^{-1}\;\mathrm{ps}$). The transfer efficiency is given by
$p_{sink}(t) = 2\Gamma_{sink}\int_{0}^t\rho_{33}(t')\mathrm{d}t'$, with
$\rho_{33}(t')$ being the population of site $3$ at time $t'$. In the inset of
Fig. \ref{fig2}, we show the behavior of $p_{sink}$ as a function of the dephasing
rate $\gamma$, at time $t \sim 10 \ \mathrm{ps}$, when one excitation is initially
injected in site $1$.

In order to describe the coupling between the FMO complex and a short laser
pulse, typically used in the laboratory to irradiate it \cite{fleming07a,fleming07b,Prokhorenko,adolphs06}, we add also a semiclassical
time-dependent Hamiltonian term, $H_{FMO-laser}(t)$, which in rotating wave
approximation takes the form
 \be
 H_{las}(t) = - \sum_{i=1}^{7} \vec{\mu}_i \centerdot \vec{e} \ E(t) \ e^{-i \omega_l t} \ \sigma^{+}_i + h.c.
 \label{hlaser}
 \ee
where $\vec{\mu}_i$ is the molecular transition dipole moment of the individual
site $i$ ~\cite{tronrud}, $\vec{e}$ and $\omega_l$ are, respectively, the
polarization and the frequency of the field, and $E(t)$ is the time-dependent
electric field. In the following, we assume $E(t)$ having the form
$$
E(t) = E_0 f(t) \; ,
$$
with $E_0 = 15 \ D^{-1} \ cm^{-1} \sim 9 \ \cdot \ 10^7 \ V/m$ (where in SI units
the Debye is given by $D\sim 3.34 \ \cdot \ 10^{-30} \ C \cdot m$), and a
time-dependent modulation
\begin{equation}
f(t)=\frac{e^{-\frac{(t-t_0)^2}{2
      \sigma^2}}}{\lambda(t)}\frac{1+\sum_{k=1}^{m} A_k \sin(\nu_k t)+
  B_k \cos(\nu_k t)}{1+\sum_{k=1}^{m} |A_k|+|B_k|} \; ,
\label{ft}
\end{equation}
with a ramp factor $\lambda(t)=1+5 [e^{200 (t-T)/T}+e^{-200 t/T}]$ (such that
$f(0) \sim f(T) \sim 0$), and $A_k$, $B_k$, and $\nu_k \equiv 2 \pi k r /T $
being parameters to be optimized by using the method described below, where
$r$ is a random number, $T$ is the time at which we want to prepare, for instance,
some desired state. Moreover, we vary also the angles $\theta$ and $\phi$ of the polarization axis
$\vec{e}$, with respect to the dipole moment of site $1$, i.e.$\theta=\theta_1 +
\Delta \theta$ and $\phi=\phi_1+\Delta \phi$, with $\theta_1$ and $\phi_1$
describing the orientation of the site-$1$ dipole moment, and $\Delta \theta$ and
$\Delta \phi$ being some free parameters.
The dipole moments of the $7$ FMO chromophores along the three reference axes are (in Debye $D$, where $1 \ D \sim 3.34 \ \cdot \ 10^{-30} \ C \cdot m$):
\begin{center}
\begin{tabular}{|c|c|c|c|}
  \hline
BChl &    X    &   Y           &    Z \\
  \hline
1    & -3.081  & 2.119  & -1.669 \\
  \hline
2    & -3.481  & -2.083 & -0.190 \\
  \hline
3    & -0.819  & -3.972 & -0.331 \\
  \hline
4    & -3.390  & 2.111  & -1.080 \\
  \hline
5    & -3.196  & -2.361 &  0.7920 \\
  \hline
6    & -0.621  & 3.636  &  1.882 \\
  \hline
7    & -1.619  & 2.850  & -2.584 \\
  \hline
\end{tabular}
\end{center}
In the following, we choose $m=7$ and $T=250 \ \mathrm{fs}$ and also chose the
carrier frequency $\omega_l$ as an additional free parameter in the optimization.
We have also worked with higher values of $m$, up to $m=25$, which resulted in
small fidelity enhancements (a few $\%$) but resulted in longer optimization
times.

As direct single site addressing is not possible in the FMO complex due to the
strongly overlapping lines we will apply quantum optimal control tools as described
more carefully in the next section to shape the laser pulse, i.e. $E(t)$, and prepare
the system in a desired physical state and control its transport dynamics. Let
us point out already here, that the parametrization of the laser pulse has been
chosen to include some additional constraints, related to typical constraints that
can be expected in future experiment in this direction. In particular, we limit
the laser power both due to experimental reasons and in order to avoid damage to
the sample or cause strong saturation. We also impose a limit on the time resolution
of the laser pulse, by means of a limit to the spectral width of the pulse, to
ensure that it does not exceed $10 \ \mathrm{fs}$, since faster modulation of
the laser is difficult to control experimentally.
\section{Optimal control: Background and Method}

Coherent control of exciton states in light-harvesting systems was demonstrated
experimentally in~\cite{herek} and theoretically in~\cite{brueggemann}. In
Ref.~\cite{herek} feedback-optimized femtosecond pulses are applied to the LH2
antenna complex from Rhodopseudomonas acidopila and to a bioinspired artificial
dyad molecule, in order to control the efficiency of the light-harvesting
dynamics. Specifically, they optimize the branching ratio of energy transfer
between intra- and intermolecular channels in the complex's donor-acceptor system
and obtain an enhancement of about $30\%$ in the LH2 system and about $10\%$ in
the artificial dyad molecule, by shaping the pulses employing feedback in an
iterative learning loop scheme.

In all mentioned experimental demonstration of control of photochemical and photobiological processes~\cite{herek,shapiro,coh-review,pierce,kosloff,rice,may},
closed-loop optimization by evolutionary algorithm
was applied~\cite{judson}. The procedure consists of
three basic components: 1) a pulse shaper, generating the
pulse shape to be tested, 2) the experiment, generating the feedback
signal by pump-probe spectroscopy, 3) a computer, running the learning
algorithm and driving the optimization. Hence, the closed-loop
optimization proceeds along the following steps: i) a random guess of a set
of pulses is shaped through the pulse shaper and then tested on the
sample; ii) then the feedback signal is evaluated and used to start an
evolutionary genetic algorithm (based on selection of `parents',
`mutations', `recombination', and `generation' of new sets of pulse
shapes); iii) a new set of pulses is obtained through the pulse shaper
and applied to the sample. These steps will be repeated until the
optimization has converged by following the so-called learning curve.

Despite the interesting applications of this technique, closed-loop optimization
tends to be effective only for population control, while coherent control
experiments cannot be performed because of the inherent shortcoming of transient
absorption (TA) spectroscopy. One way to overcome this issue could be to use 2D
electronic spectroscopy in order to get information about
the phase, which is necessary as a feedback signal in coherent control
experiments.
Here, we solve this issue, using open-loop control approach:
one first numerically optimizes laser pulses via numerical simulations
and the applies them to the sample obtaining the desired result, for
example, the experimental preparation of the sample in some
desired state with very high fidelity.

The main advantages of the open-loop approach with respect to the
closed-loop approach are two-folds. In the latter, the pulses are often
very complex, highly structured, and very demanding to interpret:
it is usually rather difficult to understand the real physical effect of
such series of consecutive pulses on the system. This limits
the understanding of the physical processes underlying a certain
biological behaviour. Moreover, repeated closed-loop experiment
rarely result in the same genetic algorithm-driven learning curve,
increasing the difficulties of the analysis of the final optimal
series and of the error estimation.
On the other hand, in the open-loop scheme the optimally-shaped
pulse is well determined and can be applied on the sample repeatedly
to increase the signal to noise level and to compare the output feedback
when changing the applied phase. It is then easier to find the explanation
for the response of the system and the physical mechanism behind it. This
problem becomes even simpler when the CRAB optimization is used as, as
explained below, it results in optimal but very simple, robust and structured
pulses.
Let us stress however that the open loop technique can be applied directly only
when the system parameters (e.g., Hamiltonian and environmental noise) are
sufficiently well known and this, indeed, makes the closed-loop approach more
feasible for several biological systems in which this information is not yet
accessible.
Here recently developed methods for quantum process tomography applied to multi-chromophoric systems, providing the decoherence of the system, the
density matrix, as well as the Hamiltonian parameters \cite{alan2011} will
be able to assist our open-loop approach. This is even more so as the robustness
of the pulse shapes that we obtain from our open-loop technique allow us to use
open loop scheme even if the system details are not well measured. Quite reasonably,
the best scheme would be a combination of these two approaches. Indeed, one could
use, for instance, closed-loop control as a means to obtain information about system
parameters, e.g. if one optimizes the pulse in the experiment then one can chose this
pulse and find out theoretically for which system parameters it reproduces the
experimental data. Then, varying the target that is to be optimized, we obtain different pulses and can repeat this investigation; each time one obtains useful information about the system parameters. Working out such a programme might be useful as it might allow an alternative to tomography. Once we apply this procedure, by using the obtained information about the system, an open loop control can be then successfully applied.

The first work where open-loop optimal control in FMO complexes is
reported in~\cite{brueggemann}. The authors investigate the control of
exciton dynamics in FMO complex, by using polarized-shaped pulses
optimized by means of a derivative functional equation for the target
function. In particular, they use shaped pulses to optimize the
excitation energy localization in a single chromophore of the FMO
complex (site $7$).

Here we extend and improve this result using the recently developed CRAB
optimization technique (described below) targeting different initial states,
to investigate different transport and decoherence processes in FMO.
Moreover, we are able to consider faster processes that are more robust against
decoherence as we optimize pulses of a few hundred femtosecond length. This allows us also
to perform coherent control, i.e. preparation of a coherent superposition state, as well excitation energy localization. Concerning the case of single site preparation, it is not feasbile to perform a clear comparison with the results of Ref.~\cite{brueggemann}, since they studied the FMO antenna complex present in the  \textit{Chlorobium Tedidum} bacterium, while here we investigate the FMO complex found in a slightly different bacterium called \textit{Prosthecochloris aestuarii}. However, from a general perspectives, it seems that our approach allows us to get much higher fidelity, and, more importantly, to use sensibly shorter laser pulse lengths ($250 \ \mathrm{fs}$, while their optimal pulse are $600 \ \mathrm{fs}$ long). These difference appear to be even more relevant and crucial when one wants to prepare a coherent state in presence of strong dephasing noise.

It should be noted that, since we are considering short pulse durations,
neglecting the double-exciton states is a good approximation, also according to
the theoretical results in Ref. ~\cite{brueggemann}. Besides, the efficiency
of the CRAB algorithm allow us to investigate more deeply the issue of
random orientations of the FMOs in the sample by considering very large samples
($10^4$ FMOs), while in Ref.~\cite{brueggemann}, due to computational limitations,
only some preliminary results were reported. In fact, we will demonstrate that even
a partial orientation of the samples by means of an external field combined with
optimal pulses will improve experimental results significantly. Finally, we show
also how to optimize the probe of the system to improve the experimental results
even more.

To achieve all this here, we use the Chopped RAndom Basis (CRAB) optimization,
introduced in Ref.~\cite{DCM}, to optimize a specific figure of merit,
e.g. population in some localized or delocalized state at
some final time or the final fidelity with respect to a
target state $\mathcal{F}(T)$, by varying the control
field entering into the Hamiltonian term in Eq.~(\ref{hlaser}).
Introducing the control field parametrization given in Eq.~(\ref{ft}) the
functional becomes a multivariable cost-function $\mathcal{F}(\Delta
\theta,\Delta \phi,\omega_l,t_0,\sigma,\{A_k\},\{B_k\})$
on which any standard minimization method can be applied.
We started with $\mathcal{O}(10^3)$ different initial random
configurations and applied a direct search algorithm, which does
not compute gradient nor Hessian,
to find the function minimum~\cite{numrec}. To minimize an
$M$-variable function, the Nelder-Mead algorithm
starts defining a $M+1$ dimensional polytope and then, in its
simplest implementation, moves it
replacing the worst point with a point reflected through the barycenter
of the other $M$ points, resulting in a
(local) minimization of the function. We used the Subplex variant
of the Nelder-Mead algorithm, which applies the same algorithm to
different subspaces to improve the convergence~\cite{rowan}.
The CRAB optimization strategy introduced above allows --as we shall
show in the following-- to find the optimal pulses to extremize
the desired figure of merit: it is efficient and versatile
as it does not need any analytical solutions of the system dynamics, it does
not compute gradients and can be easily adapted to different figure
of merits. More importantly, it includes already many experimental
constraints such as the finite band-width and power of the
control pulses.
Finally, we mention that optimal control pulses are quite robust with respect to
system parameters perturbation and noise up to a few percent, as shown in the
literature in different scenarios (see e.g.~\cite{QCapplyed1,QCapplyed2}).
This robustness arise from the fact that the optimal dynamics lies in
a minima of the functional to be extremized, thus first order
perturbations vanish.
\section{Initial state preparation}
In Ref. \cite{cdchp10}, it has been found that the coherent transfer
of the electronic excitation energy in the FMO complex takes place
essentially through two different pathways: one mediated by the state
$|+\rangle$, which is shifted towards site $3$ and leads to very fast
transfer to the reaction center, and a second one, involving the sites
$5$, $6$ and $7$, which is comparatively slow because of the energy gap
with the site $3$ and because the excitation suffers many coherent
oscillations between those sites before reaching site $3$. Indeed,
the presence of dephasing noise assists the transport because, on the one hand,
it opens up a new additional pathway, i.e. incoherent tunneling between the
state $|-\rangle$ and $|+\rangle$, and, on the other hand, it partially
suppresses the transition from $|-\rangle$ to sites $5$, $6$, $7$, and also
leads to fast incoherent oscillations between those three sites before reaching
the sink.
\begin{figure}[t]
\centerline{\includegraphics[width=.48\textwidth]{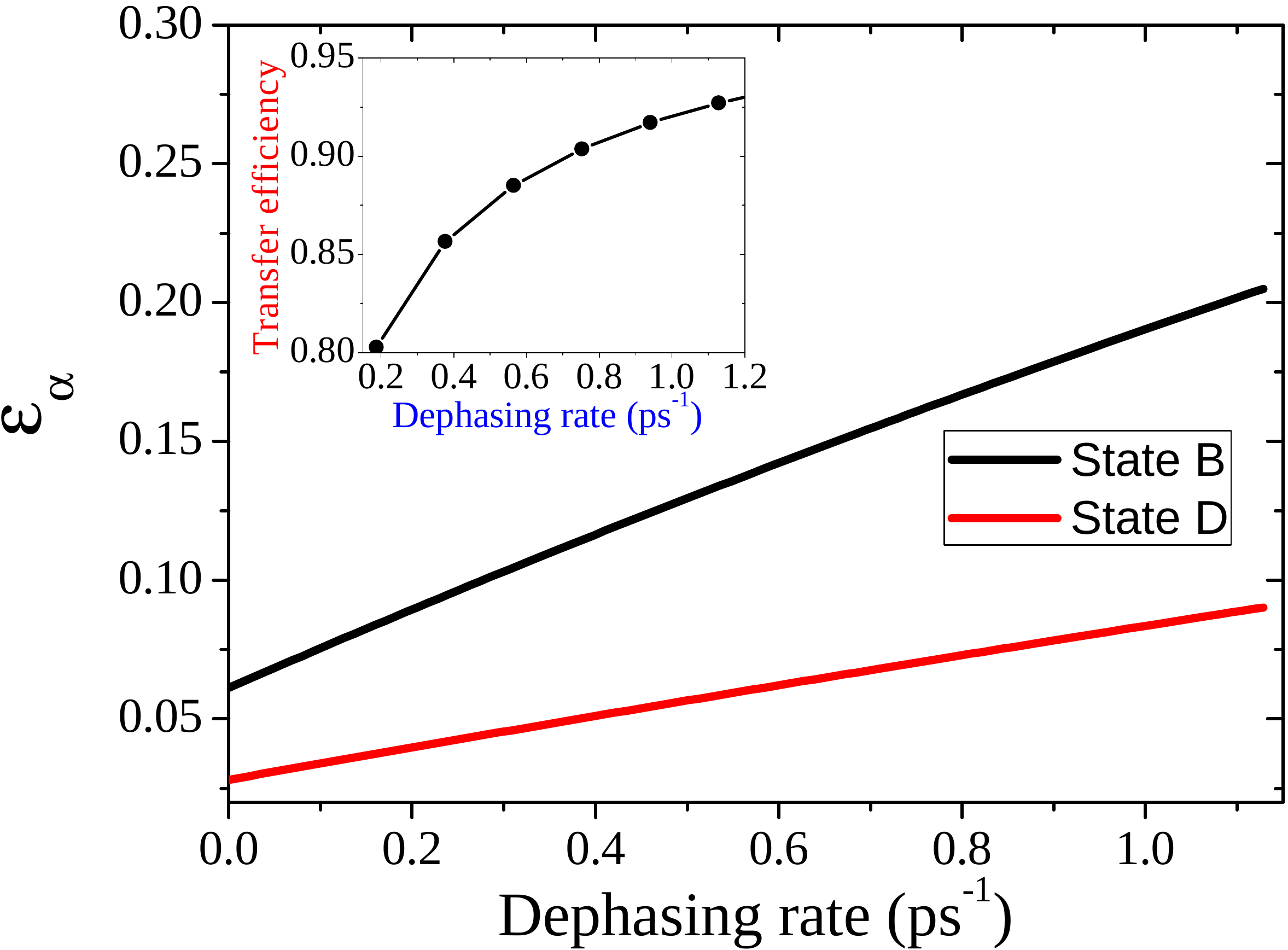}}
\caption{$\varepsilon_{\alpha}$ versus dephasing rate $\gamma$ (in units of
  $ps^{-1}$), for different prepared initial states. Specifically, our
  goal is to prepare the states $B$ and $D$ (i.e., $\alpha=B,D$). As described in the text,
these quantities are given by $\varepsilon_B = 1-\langle + | \rho | + \rangle =
1-\frac{\rho_{11}+\rho_{22}}{2}-\Re[\rho_{12}]$ and $\varepsilon_D = 1 - \rho_{5,5} - \rho_{6,6} -\rho_{7,7}$.
Inset: Transfer efficiency
  vs. dephasing rate $\gamma$ (uniform for all the sites) at $t=10
  \ \mathrm{ps}$, when one excitation is initially in site
  $1$.}\label{fig2}
\end{figure}
\begin{figure}[t]
\centerline{\includegraphics[width=.5\textwidth]{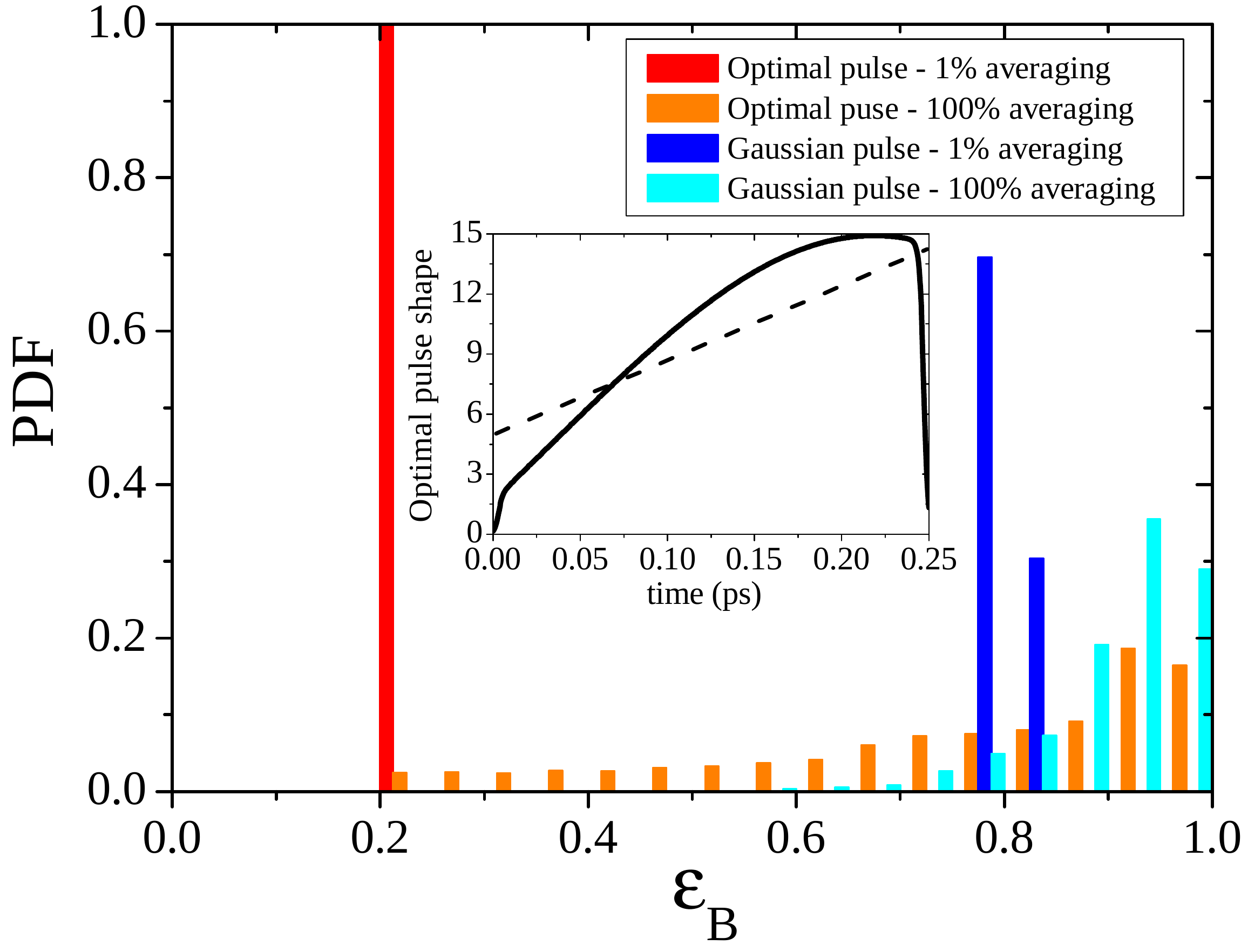}}
\caption{Probability distribution of the quantity $\varepsilon_B$
  for the preparation of the state $B$, in the presence of
  dephasing with rate $\gamma \sim 1 \ \mathrm{ps}^{-1}$, by using the
  optimal and a standard Gaussian laser pulse. We
  considered a sample of $10^4$ FMO complexes to get enough
  statistics. Inset: shape of the optimal pulse, optimized to prepare
  the state $B$ in $250 \ \mathrm{fs}$. Notice that the laser
  pulse amplitude is always shown in units of $D^{-1} \ cm^{-1} \sim 6
  \ \cdot \ 10^6 \ V/m$. Finally, the optimal values of the other free parameters are $\Delta \theta=2.5$,
  $\Delta \phi=7.6$, and $\omega_l=121.76 \ cm^{-1}$, and we find $\varepsilon_B=0.20$, without averaging.
  Interestingly enough, by considering a simpler linear shape (dashed line), we find a quite similar error, i.e. $\varepsilon_B=0.24$, by showing
  the `robustness' of the pulse shape to prepare the state $B$.}\label{fig3}
\end{figure}
Motivated by these results, we apply the CRAB optimization to find the
optimal pulses to selectively prepare a state $|D\rangle$ with maximum
probability of finding the electronic excitation in the sites $|5\rangle,
|6\rangle, |7\rangle$ (dark or non-propagating state) and
$|B\rangle \equiv |+\rangle=(|1\rangle+|2\rangle)/\sqrt{2}$
(bright or propagating state). We consider the full model for the FMO complex
(described before), but in absence of reaction center, since it is usually the
case in the current experiments on this light-harvesting system. However, the presence of the sink would not affect the state preparation process since the laser pulse is applied for a very short laser pulse. This is also confirmed by the analysis on the transport pathways discussed below.
In particular, we apply the
CRAB strategy to prepare the FMO complex (initially in the ground state) in the desired state after
$t=250 \ \mathrm{fs}$ during which we excite the system with a laser
pulse. We maximize the probability of finding the excitation in the desired
state ($B$ or $D$), that is our figure of merit, by minimizing the following
quantities
\begin{eqnarray}
    \varepsilon_B &=& 1-\langle + | \rho | + \rangle =
    1-\frac{\rho_{11}+\rho_{22}}{2}-\Re[\rho_{12}] \; ,\\
    \varepsilon_D &=& 1 - \rho_{5,5} - \rho_{6,6} -\rho_{7,7} \; .
\end{eqnarray}
In Fig.~\ref{fig2} results of the optimization are shown as a function of the
dephasing rate present in the system. As it can be seen, increasing the amount
of dephasing in the dynamics, the error $\varepsilon_{\alpha}$ increases in both
cases.

As remarked earlier, the so achieved pulses are quite robust against changes
in system parameters such as the environmental noise level. To demonstrate this
we apply the following procedure. We find the optimal laser pulse in the presence
of dephasing with rate $\gamma \sim 1 \ \mathrm{ps}^{-1}$. Then, we repeat
the optimization with different values of dephasing and find, in general, only
slightly different pulse shapes that result in values of error $\varepsilon_{\alpha}$, which differ only slightly, less than $10^{-2}$, from those obtained for the
optimal shapes we found for dephasing $\gamma \sim 1 \ \mathrm{ps}^{-1}$. In
other words, the same optimal pulse can be used to prepare some desired
state irrespective of the strength of the dephasing noise in the dynamics is.
These pulses are shown in the insets of Figs.~\ref{fig3} and \ref{fig4}. Let
us also point out that, by considering a simpler linear pulse shape (interpolating
the optimal pulse, see dashed line in the inset of Fig.~\ref{fig3}), one obtains
a very similar error, whose difference is less than $0.05$ in both cases ($D$ and
$S$ states).
\begin{figure}[t]
\centerline{\includegraphics[width=.5\textwidth]{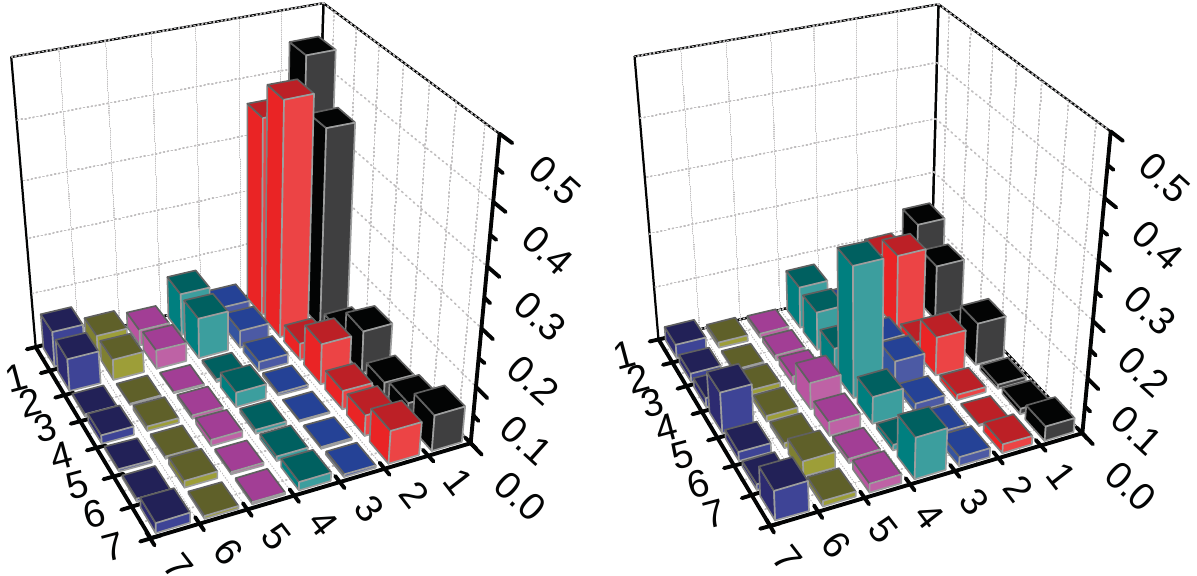}}
\caption{Modulus of the ensemble average of the elements of the FMO density matrix in the site basis (over $10^4$ samples), after applying the optimal laser pulse to prepare the system in the state $B$. We introduce also static disorder, set to $1\%$ (left) and $100\%$ (right), and we get a fidelity for the desired state equal to $89 \%$ and $49.8 \%$, respectively.}\label{fig13}
\end{figure}
\begin{figure}[t]
\centerline{\includegraphics[width=.5\textwidth]{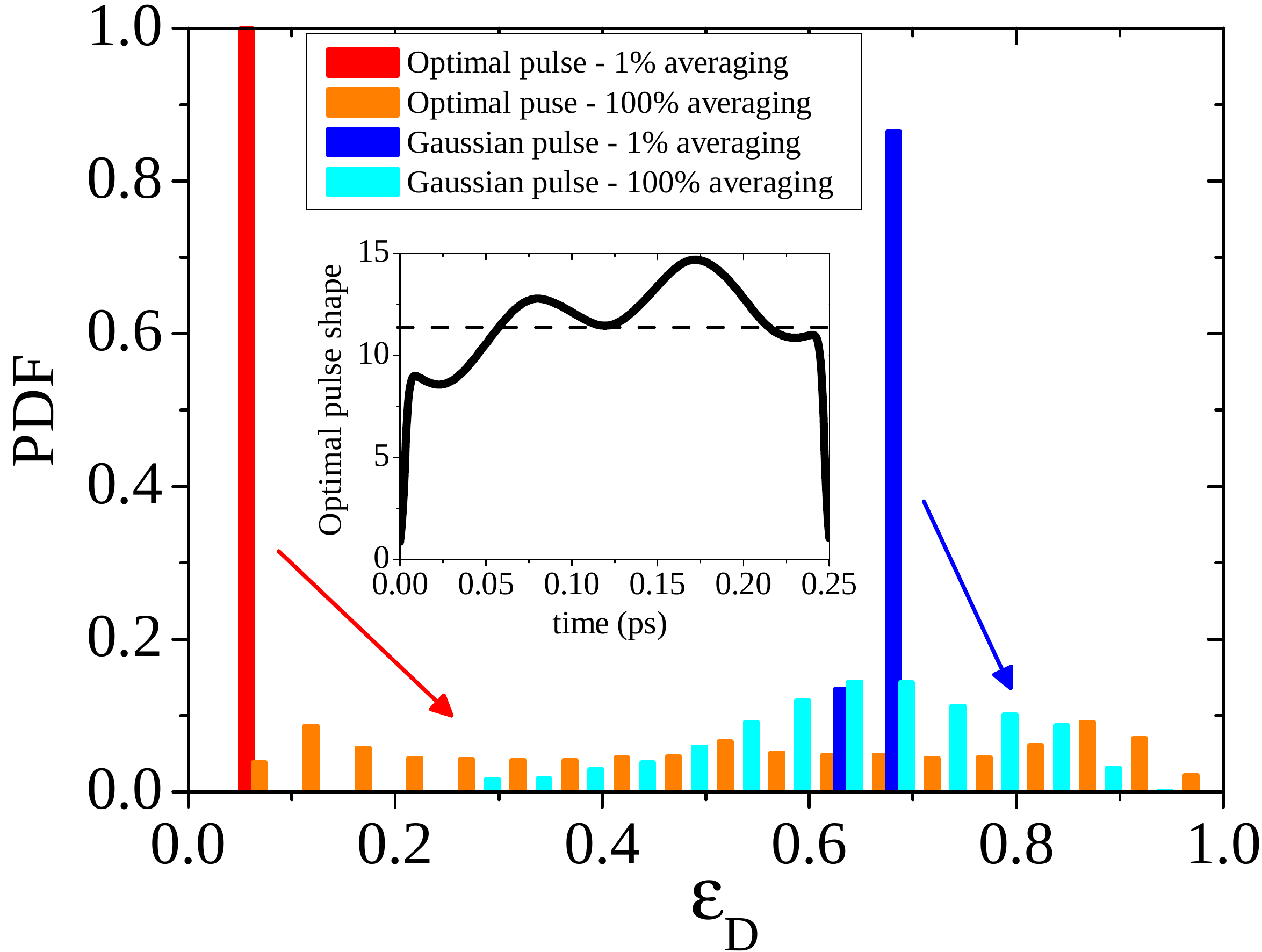}}
\caption{Probability distribution of the quantity $\varepsilon_D$
  for the preparation of the state $D$, in the presence of dephasing with rate
  $\gamma \sim 1 \ \mathrm{ps}^{-1}$, by using the optimal and a
  standard Gaussian laser pulse. As before, since in the lab one has a
  sample of FMO complexes in random orientations, we plot these pdfs
  for two extreme cases in which we add $1\%$ and $100\%$ of random
  disorder to the two angles defining the orientation of the FMO
  complex. Inset: shape of the corresponding optimal
  pulse. The optimal values of the other free parameters are $\Delta
  \theta=3.09$,
  $\Delta \phi=3.76$, and $\omega_l=504.46 \ cm^{-1}$, and we find
  $\varepsilon_D=0.09$, without averaging.
  By considering a much simpler constant laser field (dashed line),
  one still gets a very good preparation of the state $D$,
  i.e. $\varepsilon_D=0.10$.}\label{fig4}
\end{figure}
As the optimal pulse shapes depend only marginally on the noise level, the results
in Fig. \ref{fig2} can become a tool to obtain information about the noise strength
in the FMO dynamics by measuring the quantity $\varepsilon_{\alpha}$.
Notice also that, removing the experimental constraints on the pulse shape -- namely
the limit on the maximal pulse intensity and the time resolution -- it is possible
to find near-unit fidelity for any state and any value of the dephasing rate.
For instance, if one allows twice as high field strength, an improvement of even
several $\%$ for the fidelity can be obtained.
\begin{figure}[b!]
\centerline{\includegraphics[width=.5\textwidth]{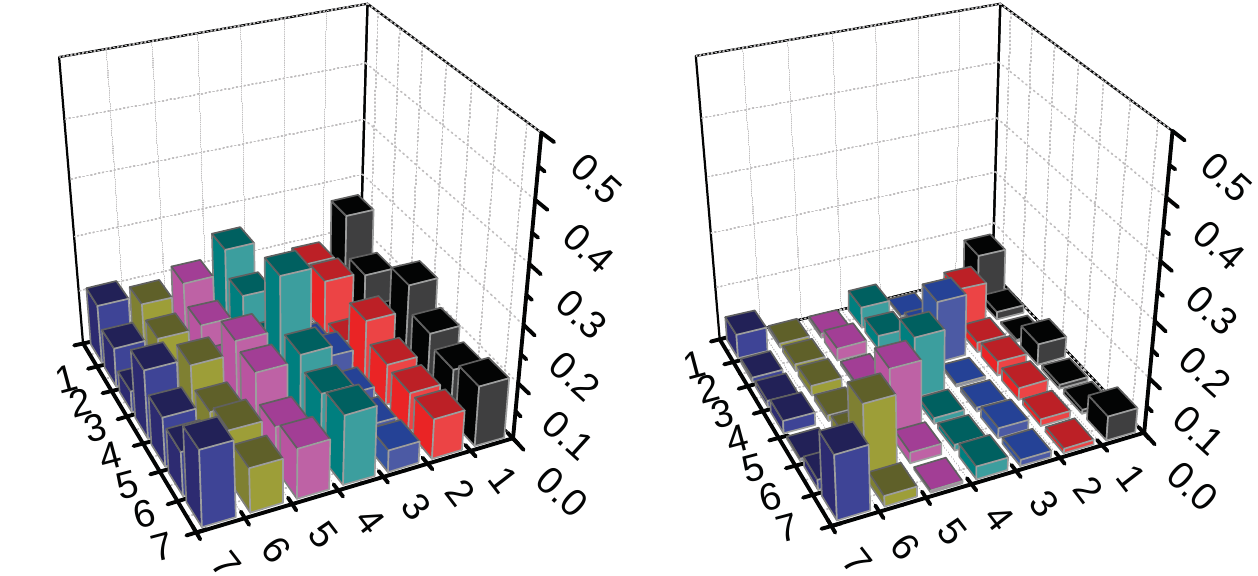}}
\caption{Modulus of the ensemble average of the elements of the FMO density matrix in the site basis (over $10^4$ samples), after applying a standard Gaussian pulse. We consider also the presence of static disorder, which is set to $1\%$ (left) and $100\%$ (right), respectively.}\label{fig14}
\end{figure}
\begin{figure}[t]
\centerline{\includegraphics[width=.5\textwidth]{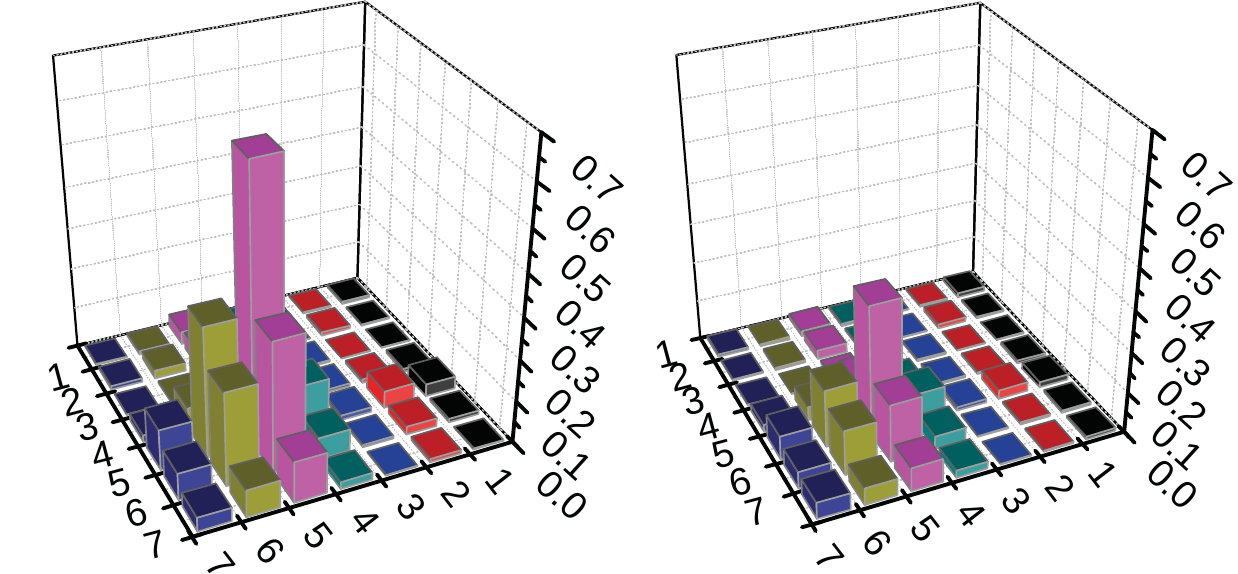}}
\caption{Modulus of the ensemble average of the elements of the FMO density matrix in the site basis (over $10^4$ samples), after applying the optimal laser pulse to prepare the system in the state $D$. Again, the static disorder is set to $1\%$ (left) and $100\%$ (right) and we get a fidelity for the desired state equal to $86 \%$ and $63.1 \%$, respectively.}\label{fig15}
\end{figure}

In realistic samples, the orientation of the FMO complexes does exhibit
significant disorder. Hence it is crucial to consider the effect of the effect
of random orientation of the FMO complexes which leads to an ensemble average
over the random distributions. In particular, we consider two extreme cases in
which we add $1\%$ and $100\%$ of random disorder, i.e. $\theta=\theta_{opt}+\eta s_1$
and $\phi=\phi_{opt}+\eta s_2$, where $\theta_{opt}$ and $\phi_{opt}$ are the
optimal values, $s_{1,2}$ are random numbers in the range $[0,2\pi]$ and
$\eta=0.01, 1$ respectively. As illustrated by the results in Figs.~\ref{fig3}
and \ref{fig4}, the corresponding distributions are fairly distinguishable. Indeed,
for the state $B$ analyzed in Fig. \ref{fig3}, the ensemble averages of $\varepsilon_B$ are
$0.207$ and $0.751$, respectively for $1\%$ and $100\%$ of orientation disorder, in the case of the optimal pulse, while they are
$0.793$ and $0.904$ in the Gaussian one, respectively. On the other hand, for the state $D$ analyzed in Fig. \ref{fig4}, the distribution averages are, respectively, $0.091$ and $0.531$ in the case of the optimal pulse, and $0.663$ and $0.634$ in the Gaussian one. Moreover, the modulus of the ensemble average of the elements of FMO density matrix in the site basis, after the state preparation by means of the optimal pulse and the Gaussian one, are shown in all the analyzed cases (in presence of static disorder) in Figs. \ref{fig13}, \ref{fig14}, \ref{fig15}.
Furthermore, we measure the fidelity $F(\rho,\sigma)$ between the desired state $\sigma$ and the one $\rho$ achieved by optimal control as $F(\rho,\sigma)=\tr[\sqrt{\sqrt{\sigma} \rho \sqrt{\sigma}}]$. Notice that, in the case of state $D$, since the quantity $\varepsilon_D$ is defined only in terms of the total population in site $5$, $6$ and $7$, in order to calculate the fidelity, we need to specify a particular goal state $\sigma$. For simplicity, we choose a target state with $70\%$ population in site $5$, $25\%$ in site $6$, $5\%$ in site $7$, and vanishing coherences. This distribution of populations is similar to the one obtained when minimizing the quantity $\varepsilon_D$. In Fig. \ref{fig13}, we notice that, if one will be able, in the lab, to orientate very well the FMO complexes in the sample, the state $B$ is prepared with very high fidelity. Additionally, when the FMO complexes are completely randomly oriented in the sample, the fidelity is much lower. Indeed, one observes a large amount of population in site $4$ which is almost in resonance with the site $1$ and $2$. However, this does not affect the transport pathway discrimination (as shown below), because the population in site $4$ goes quickly to the site $3$ as well as the state $B$. A similar behaviour is observed for the preparation of the state $D$. Nevertheless, in this case the high values of the fidelity are more robust to the introduction of orientation disorder, since one wants only to populate the high-energy sites $5$, $6$ and $7$ neglecting any control of the phase terms. Instead, in Fig. \ref{fig14}, one clearly observes that a gaussian laser pulse is not able to selectively prepare specific states. Furthermore, when $100 \%$ of orientation disorder is considered, the gaussian pulse does almost equally populate all the sites with vanishing cross-term correlations (completely mixed state).
Finally, it is worth pointing out that an experimental partial orientation of the sample will be able to sensibly enhance these fidelities, as shown below in the section about the orientation.
In the following, we will use these optimal laser pulses, found to
prepare those selected states, in order to investigate the different transport
pathways in the FMO complex dynamics.
\section{State preparation dependent transport}
\begin{figure}[t]
\centerline{\includegraphics[width=.5\textwidth]{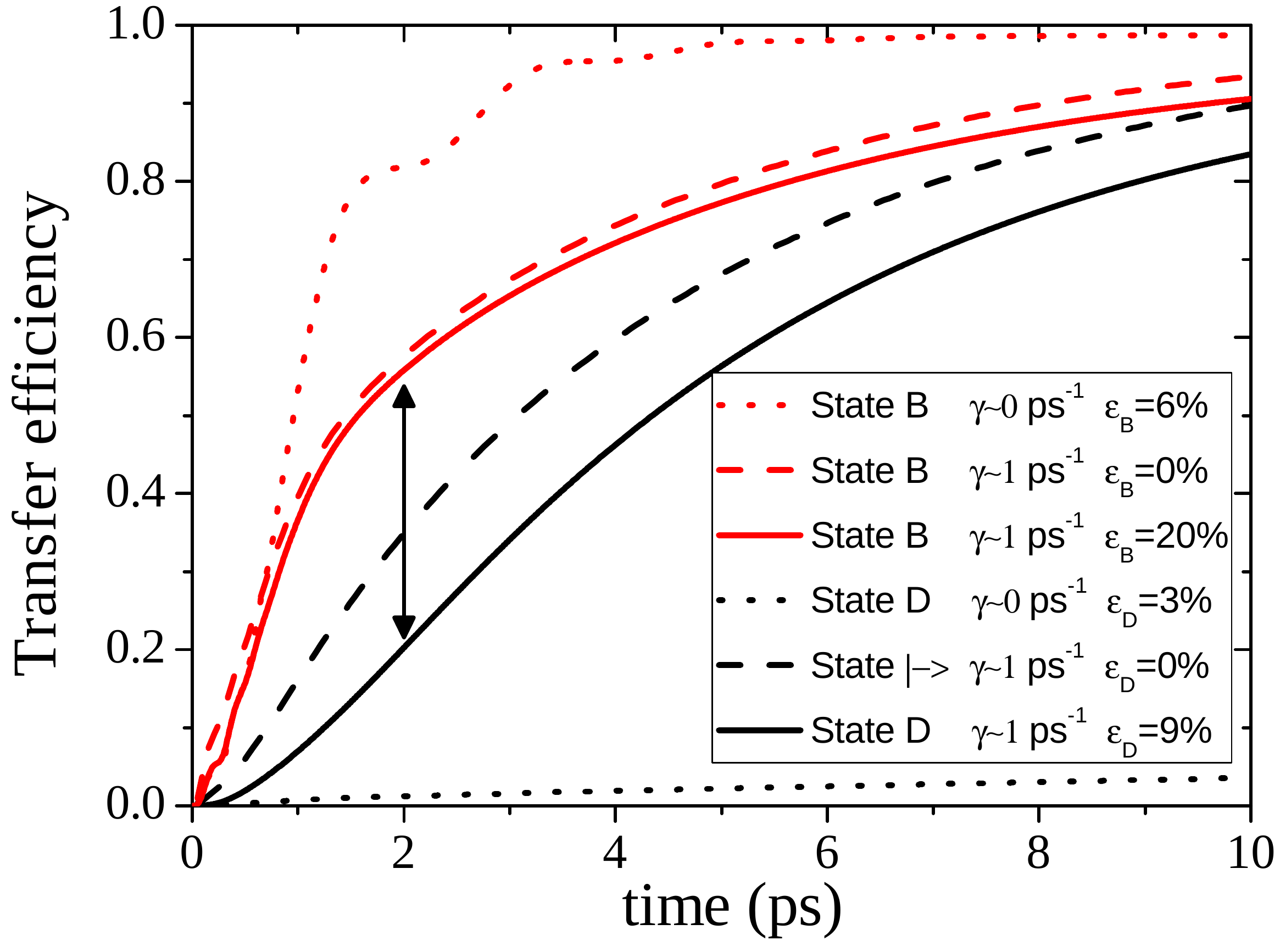}}
\caption{Transfer efficiency as a function of time ($\mathrm{ps}$) for
different initial states, i.e. $|D\rangle$, $|-\rangle$ and $|B\rangle$,
for two different values of dephasing rate $\gamma=0$ and $\gamma \sim 1
\ \mathrm{ps}^{-1}$. Moreover, in one case, we consider also an idealized
preparation of the state, i.e. $\varepsilon_{\alpha}=0$. In the other
cases, we consider the more realistic scenario when in the lab one
prepares similar states applying the optimal laser pulses, in absence
and presence of dephasing noise.
}\label{fig5}
\end{figure}
We now compare the transfer efficiency into the reaction center
corresponding to the two different initial states that we have
introduced before, including the
Lindbladian term in Eq. (\ref{sink}) -- see Fig. \ref{fig5}. Notice that now the sink is included in the dynamics from the beginning, even if the optimal pulses were obtained without sink. Indeed, as discussed also above, it does not sensibly affect the state preparation since the pulse is applied for a very short time scale compared to the transfer rate from the site $3$ to the reaction center.

In the absence of dephasing noise, as expected from the results in
Ref. \cite{cdchp10}, the population initially prepared in sites $5$,
$6$ and $7$ is basically trapped and only very slowly gets into the
reaction center, while the state $|+\rangle$ is almost in resonance
with the site $3$ and the transfer rate to the sink is much
faster. However, increasing the amount of dephasing, these
destructive interference effects are reduced, destroying the so-called invariant
subspaces or dark states (see Ref. \cite{ccdhp09} for more details), and the efficiency
discrepancy between the fast and the slow pathways decreases, as shown
in Fig. \ref{fig5}. Particularly, the ratio fast/slow pathway transfer
efficiency is about $2.5$ for $\gamma \sim 1 \ \mathrm{ps}^{-1}$ and
about $80$ for $\gamma=0$, at time $t=2 \ \mathrm{ps}$.
These results are particularly robust against various possible
experimental inaccuracies: In Fig. \ref{fig5} we show that,
although the state preparation by the optimal laser pulse is
not perfect, the corresponding behaviour is still good enough to distinguish
the slow and fast transport pathways, compared to the case
in which the specific states are exactly set in the numerical simulation.
\begin{figure}[t]
\centerline{\includegraphics[width=.5\textwidth]{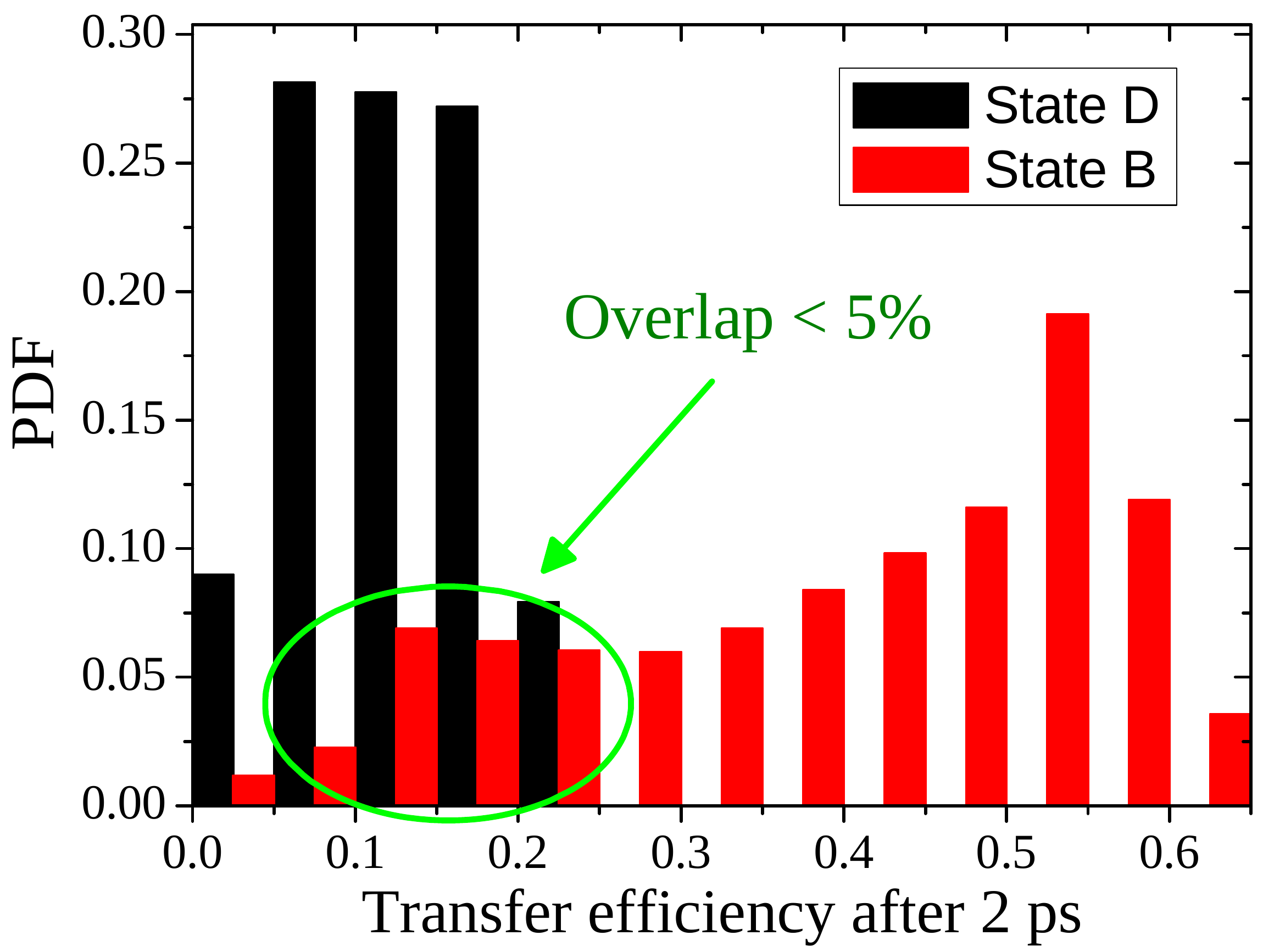}}
\caption{Probability distribution of the transfer efficiency into the sink
after $2 \ \mathrm{ps}$, in the presence of local dephasing with rate
  $\gamma \sim 1 \ \mathrm{ps}^{-1}$, for the optimal preparations of the state $D$ and the state $B$, with $100\%$ of random
disorder in the orientation of $10^4$ FMO complexes. The two
distributions are distinguishable with less
than $5\%$ of error.}
\label{fig6}
\end{figure}
In other words, it turns out that, even if the state preparation is
not perfect, one finds a very similar
behavior for the transfer efficiency. This fact is fundamental
to reproduce these results in the laboratory
since the experimental fidelities will be smaller than the theoretical
ones.
Finally, we also consider the realistic experimental scenario in which
a very large ensemble of FMO molecules is studied simultaneously in the lab
with different random orientations. It turns out that, even
with $100\%$ of random disorder in the orientations of the FMO
molecules, the difference between the transfer efficiency
after $2 \mathrm{ps}$ of the `dark' and `bright' states is still measurable.
Indeed, as shown in Fig.~\ref{fig6}, the two
distributions of transfer efficiency are distinguishable with less
than $5\%$ of error. The ensemble averages are around $0.12$ in the case of the state $D$, and $0.39$ for the state $B$.
In conclusion, the proposed analysis appears to
be observable in a real experiment.
\begin{figure}[t]
\centerline{\includegraphics[width=.5\textwidth]{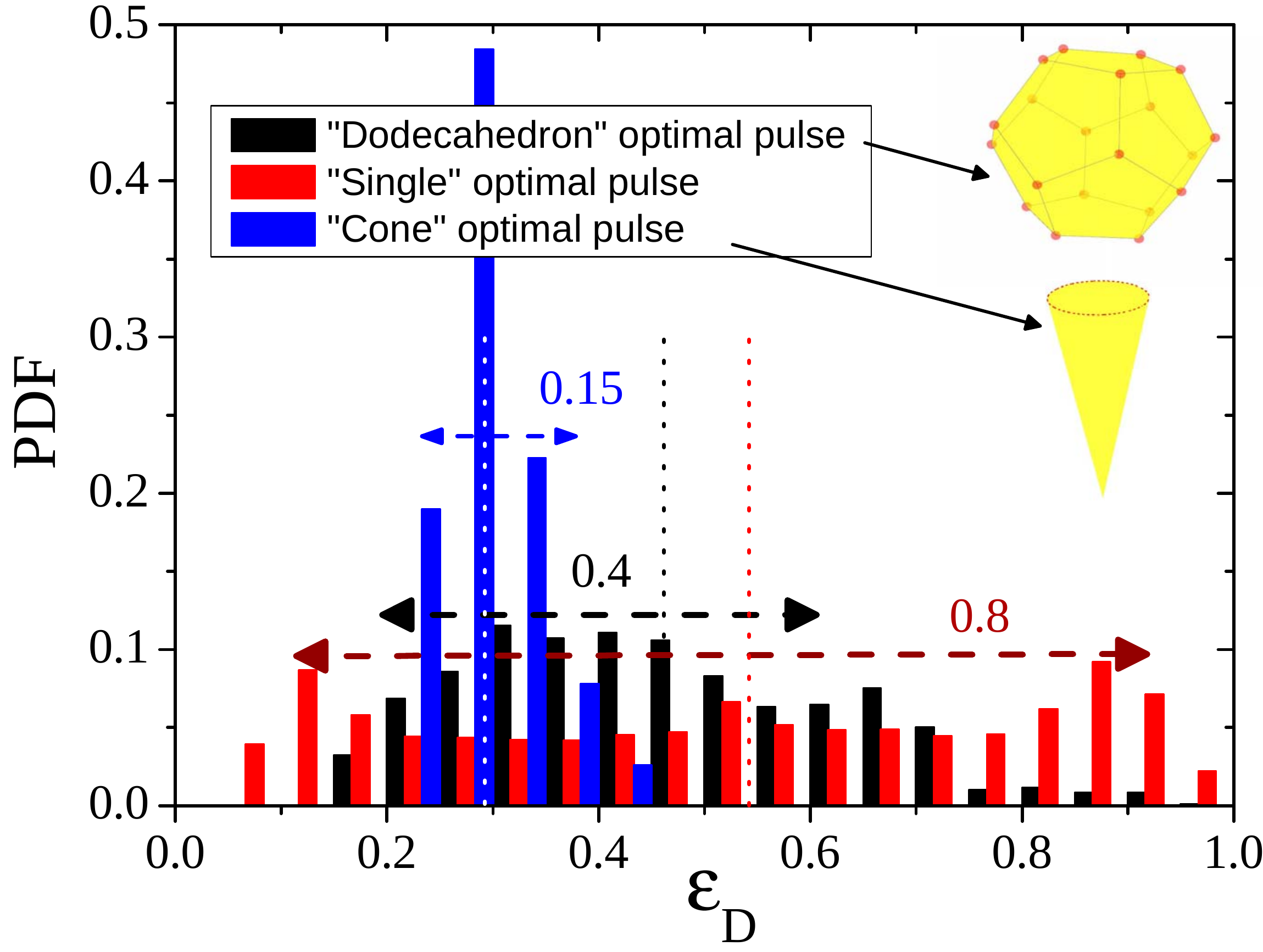}}
\caption{Probability distribution of the quantity $\varepsilon_D$
for the preparation of the state $D$, in the presence of dephasing with rate
$\gamma \sim 1 \ \mathrm{ps}^{-1}$, by using the optimal pulses
obtained by averaging on a single FMO complex with a specific
orientation (as in Fig. \ref{fig4}), by averaging on a small sample of
$20$ (almost isotropic) FMO orientations pointing the $20$ vertices of
a dodecahedron, and on a sample of $21$ orientations inside a cone
with an opening angle of $0.1 \ \pi$. Hence, one consider the
proability distribution of
$\varepsilon_D$ when these optimal pulses are applied to a sample of
$10^4$ FMO complexes in completely random orientations in the first
two cases and, for the last case, randomly oriented inside that
cone. The shape of the corresponding optimal pulses are similar to the
one in the inset of Fig. \ref{fig3}. Moreover, we find, for the
dodecahedron case, $\Delta \theta=2.66$, $\Delta \phi=2.61$,
$\omega_l=277.10 \ cm^{-1}$, while, for the cone case, $\Delta
\theta=-2.21$, $\Delta \phi=1.86$, $\omega_l=274.87 \ cm^{-1}$. The
dashed lines represent the corresponding averaged values,
i.e. $0.53$ (single orientation), $0.46$ (dodecahedron), $0.29$
(cone). The corresponding distribution widths are also plotted.}
\label{fig7}
\end{figure}
\section{Optimal control pulse for experiments on differently oriented systems}
In the lab one generally has a sample of many FMO complexes with random
orientations and, hence, the laser pulse could be optimized also taking it into account.

In order to cover almost isotropically all the different orientations of the
photosynthetic system in the sample, we consider the $20$ directions pointing
to the $20$ vertices of a dodecahedron, and we optimize the pulse in order to
minimize the quantity $\varepsilon_{\alpha}$ averaged over a sample of $20$ differently oriented FMO complexes.
Notice that the computational efficiency of the CRAB algorithm
allows us to consider a much higher number of directions and we consider
here only $20$ just for simplicity and for illustration purposes. Indeed, the
time complexity of such algorithms scales linearly with the number of orientations
to average over and the optimization in the case of $20$ directions takes only a
time of the order of hours on a single standard CPU.

In Fig. \ref{fig7}, we show the probability distribution for
$\varepsilon_D$ when this new optimal pulse is applied to a sample of
$10^4$ FMOs in random orientations. As comparison, we plot also the
corresponding probability distribution when we instead apply the optimal pulse
used in Fig. \ref{fig4}. It turns out that this new pulse is somehow more
robust and provides an error distribution whose width is about one
half of the one obtained with the pulse optimized with a single FMO
system. This could suggest a way to improve the state preparation of
the FMO complex in the real experiment where one has a sample
containing many FMO molecules in a solution. However, concerning the
preparation of the more delicate state $B$ this technique does not
provide any noticable improvement. Moreover, regarding the transport
properties, the results are similar to those presented in
Fig. \ref{fig6} and the overlap of those two distributions is still
less than $5\%$ but does not decrease further. Notice that this
overlap is already rather small to be able to be experimentally
observed.
\begin{figure}[t]
\centerline{\includegraphics[width=.49\textwidth]{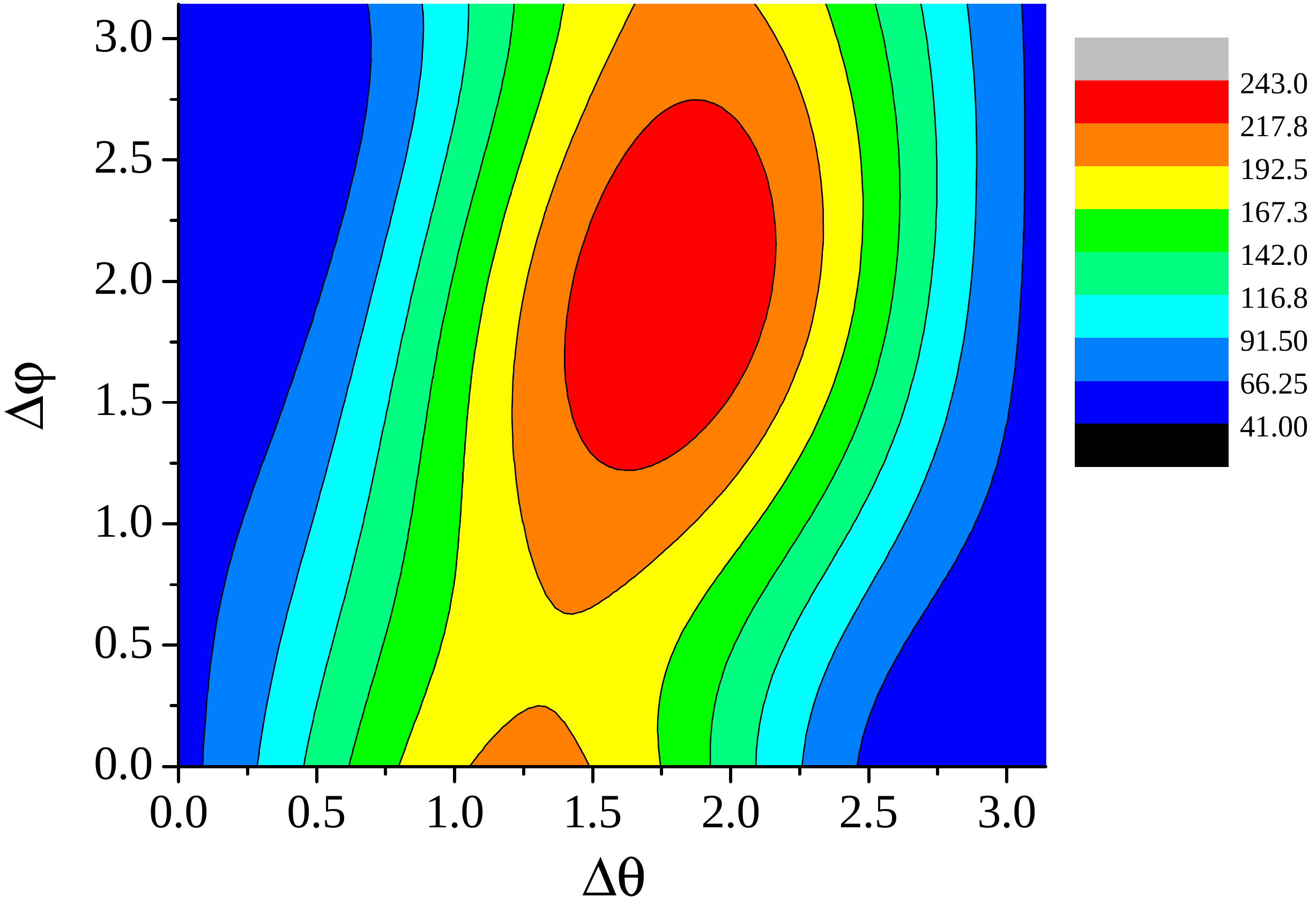}}
\centerline{\includegraphics[width=.49\textwidth]{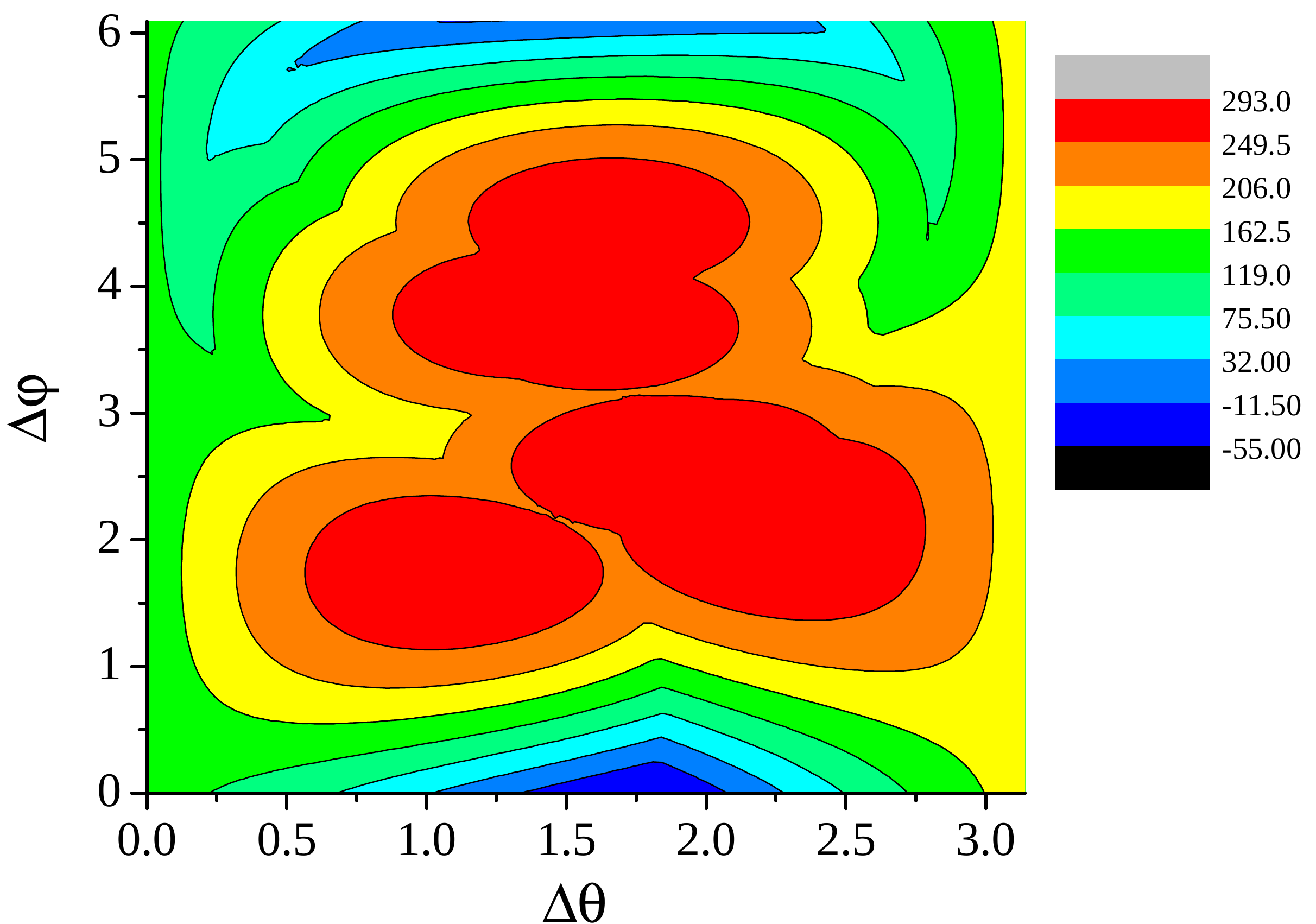}}
\caption{Top: Behaviour of the quantity $\Delta$ (defined in the
  text), in units of $cm^{-1}$ ($\sim 0.1 \ \text{meV}$) as a function
  of the angles $\Delta \theta$ and $\Delta \phi$ defining the
  orientation of the FMO complex, in the case of $\omega_l=-1000
  \ cm^{-1}$ and $E_0=70 \ D^{-1} \ cm^{-1} \sim 42 \ \cdot \ 10^7 \ V/m$. The difference between
  the maximum and the minimum value is comparable to the thermal
  energy and the maximum is obtained for $\Delta \theta \sim 1.75$ and
  $\Delta \phi \sim 2$, which seems to be the preferred orientation
  when the sample is subjected to this constant laser field.
Bottom: maximum value of the Rabi frequencies $\vec{\mu}_i \centerdot
\vec{e} E_0$ versus $\Delta \theta$ and $\Delta \phi$. Notice that the
values of the detunings are much larger than the Rabi frequencies.}
\label{fig9-10}
\end{figure}
Finally, we consider an intermediate case in which the FMO complexes can be
partially oriented along a cone-shaped orientations. In particular, we repeat
the analysis above but for $21$ directions inside a cone with a $10\%$
opening angle, i.e. $0.1 \pi$, and we optimize the pulse in order to
minimize the quantity $\varepsilon_{\alpha}$ averaged over $21$ FMO
complex evolutions. Then, we apply this optimal pulse to a sample of
$10^4$ FMOs in random orientations inside this code and we calculate
the probability distribution for $\varepsilon_D$ - see
Fig. \ref{fig7}. We find that the width of the error distribution is
further squeezed and shifted to smaller errors, i.e. higher
fidelities.
\section{Towards the orientation of the FMO sample}
\begin{figure}[t]
\centerline{\includegraphics[width=.5\textwidth]{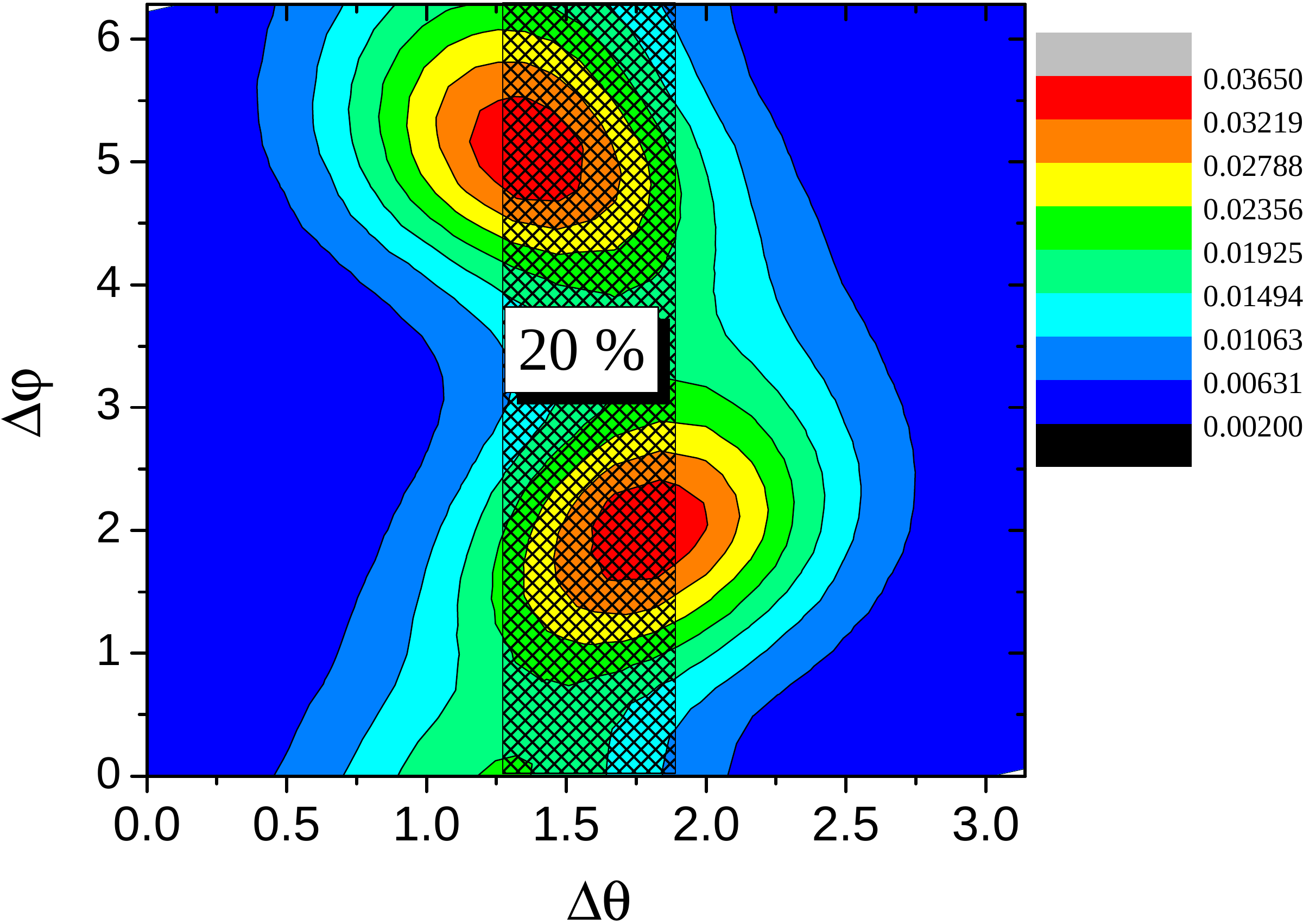}}
\caption{Boltzmann-Gibbs probability distribution in orienting the
  FMOs in the sample, as a function of the angles $\Delta \theta$ and
  $\Delta \phi$, in the case of $\omega_l=-1000 \ cm^{-1}$ and $E_0=70
  \ D^{-1} \ cm^{-1}$. The dashed area represents the region in which
  we will sum up all the probabilities in the case, for instance, of a
  $20\%$ cone opening angle, as shown in Fig. \ref{fig12}. Notice, however, that a slightly tilted strip
  would get higher probability for the same width and so further improves our results.
The Gaussian pulse shape, used as comparison in the population behavior,
is also shown (dashed line).}
\label{fig11}
\end{figure}
Here, we show how one could try to orientate the FMO systems in the experimentally available sample, by using some simple classical mechanics arguments.
Following Ref. \cite{FMO}, we model each monomer of the FMO complex as a disk, whose mass and radius can be reasonably estimated
to be equal to $M \sim 80 \ \text{kDa} \sim 15 \ 10^{-23} \text{Kg}$
and $R= 2 \ \text{nm}$ (including the protein scaffolding). The
corresponding moment of inertia (with respect to one of its diameters
as rotational axis) is hence $I=\frac{1}{4} M R^2 \sim 1.125
\ 10^{-31} \text{Kg} \ \text{m}^2$. The rotational energy is
$E=\frac{1}{2} I \omega^2$, with $\omega$ being the angular velocity
in radians per second, i.e. the derivative of the angle rotated with
respect to time $\omega=\frac{d \theta}{d t}$. At room temperature, by
neglecting the friction due to the presence of a solution and,
possibly, other more sophisticated effects, just to get an initial
rough estimation, the time $t_{rot}$ it takes for the system to rotate
by an angle $\pi/2$ (which is roughly the average angle by which a
complex has to be rotated) is trivially given by
$t_{rot}=\frac{\pi}{2} \sqrt{\frac{I}{2 E_{th}}} \sim 6 \ \mu s$, with
$E_{th} \sim 25 \ \text{meV}$ being the thermal energy. Actually,
given that the disk will carry out a sort of random walk, the rotation time could be
much longer than what is estimated here by means of this simple analysis.

Moreover, we
investigate the energy landscape as a function of the orientation of
the FMO complex, when is subjected to a constant electric field $E_0$
polarized along the axis $\vec{e}$ with carrier frequency $\omega_l$,
i.e. the following quantity
\begin{equation}
\Delta= \sum_{i=1}^{7} \frac{|\vec{\mu}_i \centerdot \vec{e} \ E_0 |^2}{\omega_i-\omega_l} \; .
\end{equation}
By varying the orientation of the FMO complex in terms of the two
angles $\theta$ and $\phi$, the energy difference between the maximum
and the minimum value is comparable to the thermal energy and one gets
the maximum for $\Delta \theta \sim 1.75$ and $\Delta \phi \sim 2$ -- see
Fig. \ref{fig9-10}. The other parameters are chosen as $\omega_l=-1000
\ \text{cm}^{-1}$ and $E_0=40 \ D^{-1} \ cm^{-1}$.
\begin{figure}[t]
\centerline{\includegraphics[width=.48\textwidth]{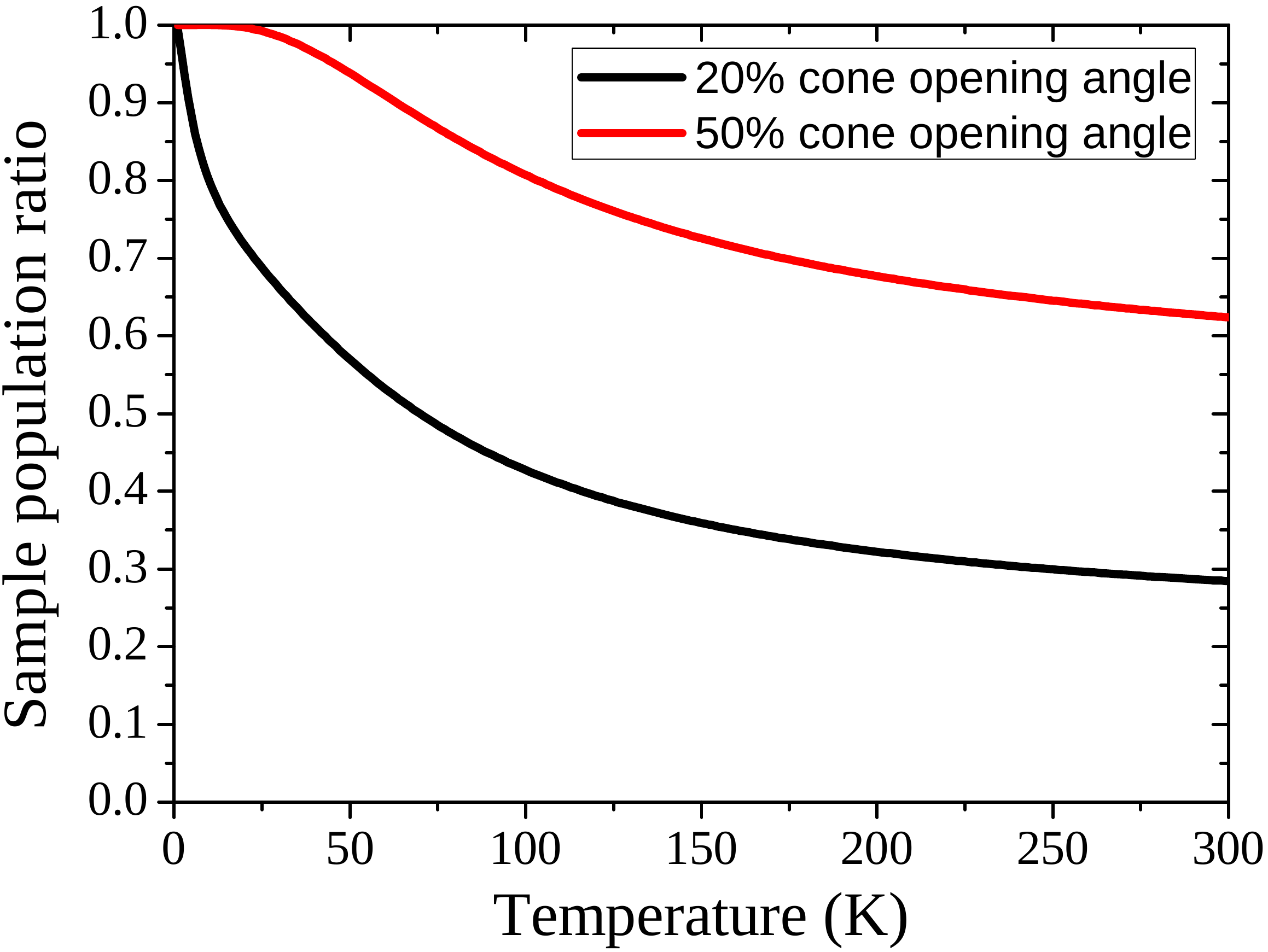}}
\caption{Sample population ratio of FMOs oriented within
  a cone with some opening angle ($20\%$, i.e. $0.2 \pi$, and $50\%$,
  i.e. $0.5 \pi$), as a function of temperature (K), with $\omega_l=-1000 \ cm^{-1}$, and $E_0=70 \ D^{-1} \ cm^{-1}$,
  according to the Boltzmann-Gibbs distribution.}
\label{fig12}
\end{figure}
\begin{figure}[b!]
\centerline{\includegraphics[width=.5\textwidth]{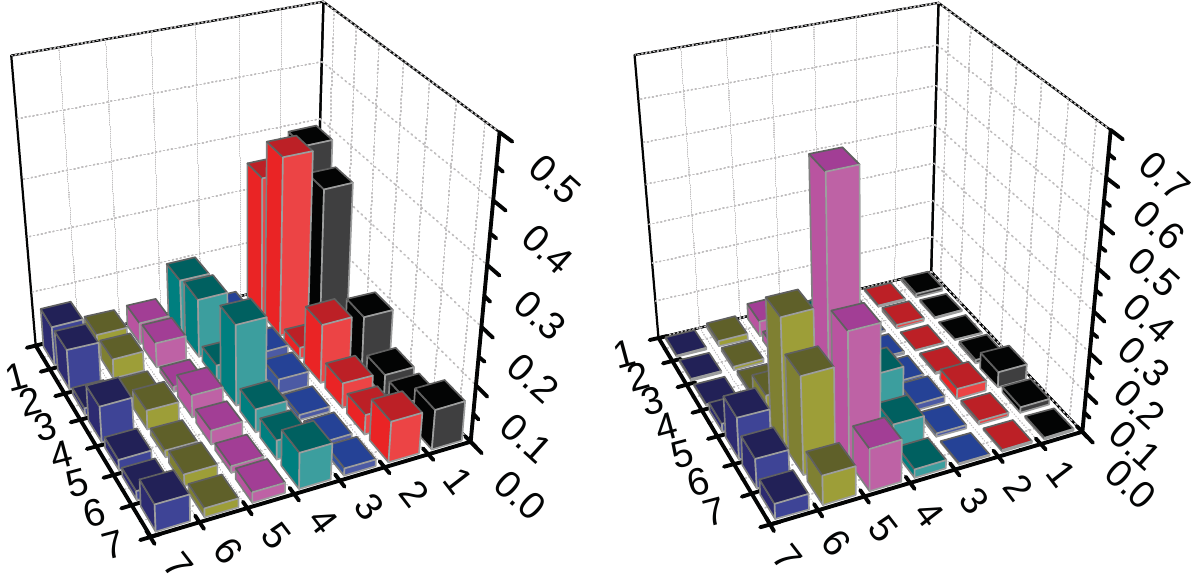}}
\caption{Modulus of the ensemble average of the elements of the FMO density matrix in the site basis (over $10^4$ samples), after applying the optimal laser pulse to prepare the system in the state $B$ (left) and $D$ (right), respectively. We introduce a static disorder, set to $10\%$ in both cases, considering that some orientation will be obtained in the experiment. The fidelity for the preparation of state $B$ and $D$ are $74.4 \%$ and $85.1 \%$, respectively.}\label{fig16}
\end{figure}
Finally, following the simple reasoning above, we explicitly calculate
how many FMO systems can be oriented around some direction at a
certain external temperature, since the thermal fluctuations will
unavoidably try to disorientate the sample. To do this, we calculate the
probability for the FMO to be oriented at a certain angle by assuming
that they follow a Boltzmann-Gibbs distribution, in the presence of a
constant electric field - see Fig. \ref{fig11}. Then, by using the
obtained probability distribution, we also evaluate how many FMO
systems are oriented within a cone with a certain opening angle as a
function of the temperature and for different values of opening
angle. The corresponding results are shown in Fig. \ref{fig12}. By
decreasing the temperature, the amount of oriented FMO systems in the
sample increases and this would give us higher fidelities in the state
preparation analysis and for the other related results above.
Therefore, it seems feasible to orientate the FMO complexes in the
sample to some extent using far detuned laser light. The typical
individual optimal control experiments take of the order of $\sim \mathrm{ps}$
which, importantly, are several orders of magnitude shorter than the
time for the FMO complexes to loose orientation after the detuned
laser beam has been switched off. Hence, we can avoid interference
between the orientation laser and the actual optimal control experiment.
Finally, let us point out that a partial orientation of the sample will allow one to sensibly improve the state preparation results shown above - see Fig. \ref{fig16}. Indeed, a net improvement is observed, when compared to the $100\%$ disorder case shown in Figs. \ref{fig13} and \ref{fig15}.
\section{Optimal probe}
In the context of controlling a molecular systems by optimal
femtosecond laser pulses, a similar approach can be reasonably used to
optimize the probe pulse absorption in a pump–probe scheme. A
successful demonstration of the optimization of the absorbance of the
probe pulse by optimal control techniques, but based on a derivative
functional equation, was shown for a prototypical molecular three
level system in Ref. \cite{kaiser}. Here, we repeat the analysis above
for a probe laser pulse applied to the FMO complex by using the
CRAB algorithm. In order to compare our
theoretical predictions, e.g. in Fig. \ref{fig6}, with the
experimental data, since there is no reaction center in the FMO
complex sample used in the lab, one has to measure the population in
site $3$ as a function of time and then calculate the corresponding
transfer efficiency, as defined in the model in Eq. (\ref{sink}). To do
that, a probe laser pulse is applied to the sample and the corresponding
absorption intensity is detected. Usually, the probe pulse is a
Gaussian laser pulse on resonance with the site whose population one
wants to measure, e.g. site $3$. Here, we apply the optimal control
tools to analyze whether a shaped pulse can detect the site
population, particularly in site $3$, with an higher `resolution', as
compared to a simpler Gaussian pulse. Actually, it will turn out that, when one considers only
a single FMO complex, very high fidelity ($99\%$) are already obtained by a gaussian pulse oriented along some optimal polarization axis and the pulse shaping will not give significant improvements. On the other side, if one has a sample of fully randomly oriented systems, both gaussian and an optimally shaped pulse bring to very low fidelities since the orientation disorder is too strong. However, if a partial orientation will be feasible from the experimental point of view, the control on the pulse shape sensibly increase the probability of successfully probing site $3$.
Specifically, we use the model
above for a single FMO complex and we consider the case where
initially all the population is in site $3$ and we want to find the
pulse which is able to detect this population by absorption, i.e. by
removing population from the site $3$. To apply the optimal control
approach as defined above, we use the following error function
\begin{eqnarray}
\varepsilon_P &=& \rho_{33} \; ,
\end{eqnarray}
with $\rho_{33}$ being the site-$3$ population.

Moreover, the pulse is
applied for a time interval of $t=125 \ \mathrm{fs}$. We find that, if
there is just one FMO complex in the sample, the optimization will
simply orientate the probe pulse polarization axis along some
optimal direction (related to the site $3$ dipole moment). Apart from the polarization control, the further optimization of the
pulse shape does not bring further significant gains. Hence a Gaussian
pulse will reliably detect the
site-$3$ population (absorption efficiency, i.e. $1-\varepsilon_P$, of
$99\%$). However, if one applied this Gaussian pulse polarized along
this optimized direction to a sample of randomly oriented FMOs, the
absorption efficiency shows an almost flat probability distribution in
the range $[0.45,0.99]$, which is actually a bit higher for smaller
efficiencies, i.e. the method does not provide us with a more
efficient absorption signal (compared to the traditional way), and has
associated an error of about $50\%$. If we apply the optimal control
algorithm to find the best probe, averaged over $20$ isotropic
orientations according to the dodecahedron above, this does not
improve sensibly the results either (data not shown).
\begin{figure}[t]
\centerline{\includegraphics[width=.48\textwidth]{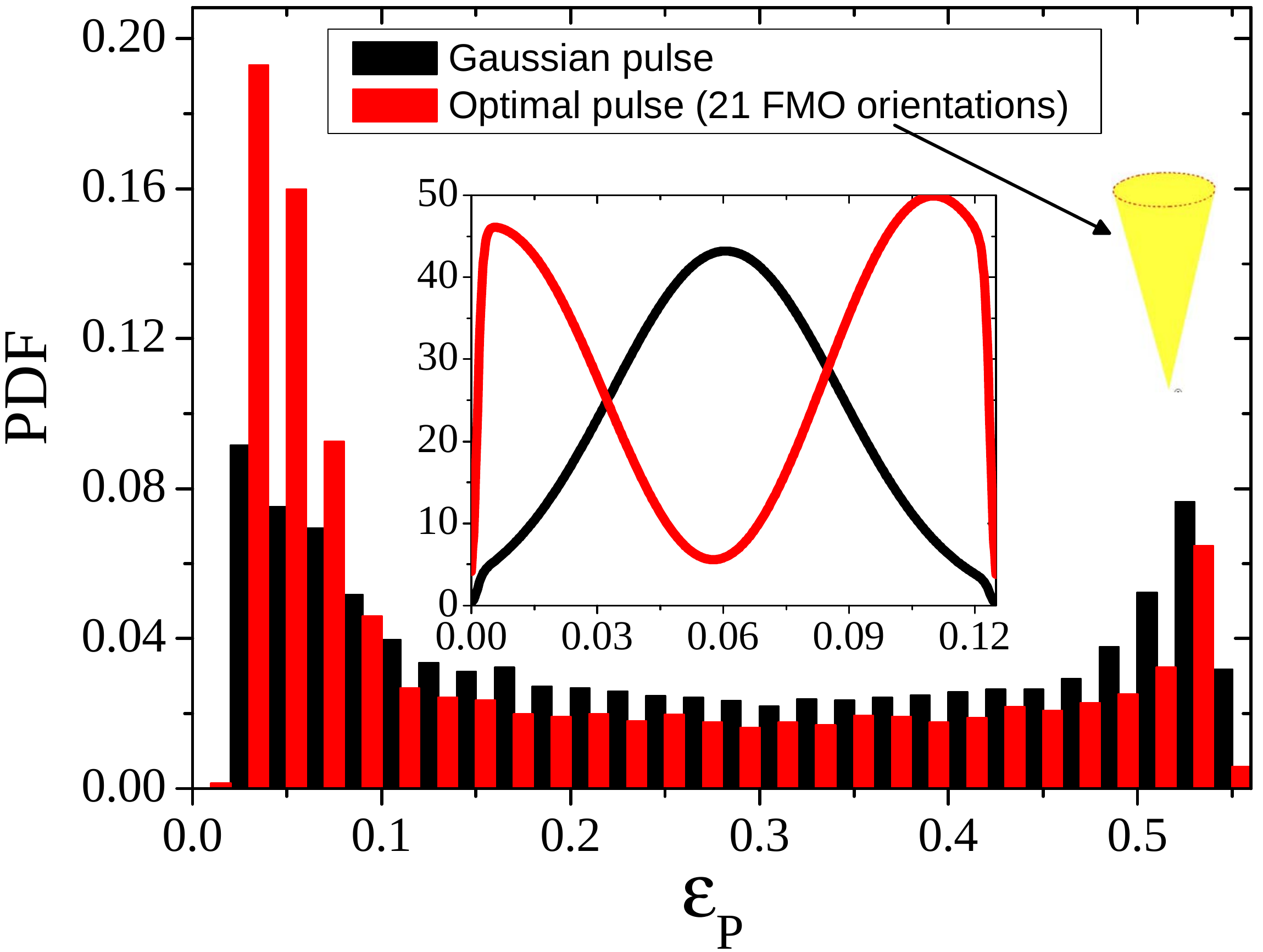}}
\caption{Probability distribution of the quantity $\varepsilon_P$,
  when the FMO is initially prepared with all the population in site
  $3$ and then subjected a laser probe pulse for a time interval of
  $125 \ \mathrm{fs}$ to detect the site-$3$ population, in the
  presence of dephasing with rate $\gamma \sim 1
  \ \mathrm{ps}^{-1}$. In particular, we consider the case of a
  Gaussian pulse on resonance with the site $3$,
  i.e. $\omega_l=0\ cm^{-1}$, and polarized along the optimal (wrt a
  single FMO) orientation, i.e. $\Delta \theta=2.92$, $\Delta
  \phi=1.85$, and the case of an optimal pulse obtained by averaging
  the error function $\varepsilon_P$ over $21$ directions inside a
  cone with an opening angle of $0.1 \ \pi$, whose optimal parameters
  are $\Delta \theta=5.61$, $\Delta \phi=1.39$, $\omega_l=12.63
  \ cm^{-1}$. To get the probability distribution, we apply both pulses to a sample of
  $10^4$ samples randomly oriented inside that cone. Inset: shape of
  both (Gaussian and optimal) probe pulses.}
\label{fig8}
\end{figure}
Finally, we consider the case in which the sample is partially
oriented, particularly within a cone of $0.1 \pi$ opening angle, and
we optimize the pulse along $21$  orientations inside this cone, as
done above for the pump. Hence, we apply this optimally shaped
pulse to a sample of $10^4$ FMOs randomly oriented inside this cone
and we compare the absorption efficiency to the case of a Gaussian
pulse, polarized along the optimal orientation obtained for a single
FMO. We find that the optimally-shaped pulse gives a probability of
high absorption efficiency (small $\varepsilon_P$), which is more than
twice larger than the one with a Gaussian
pulse - see Fig. \ref{fig8}. Therefore, in presence of
structural disorder but achieving a partial orientation of the
FMOs in the sample, both polarization and shape
optimizations may enhance the probe pulse absorption in a pump-probe
scheme, which is crucial for an efficient discrimination of dynamical
properties, like the identification of transport paths discussed
here.
\section{Summary and prospects}
In summary, the experimental verification in bio-molecular
systems of fundamental building blocks of quantum dynamics
in presence of environment~\cite{Aspuru08,PlenioHuelga08,ccdhp09,castro08,ccdhp10,patrick,cdchp10,chp10}
requires the development of
novel experimental tools and theoretical methodology. In this
work we have contributed to this effort with the demonstration
that novel methods from the theory of optimal control
can be combined with ultra-fast laser pulses to provide enhanced
diagnostic tools suggesting promising new routes for
experiment. In particular we have introduced and applied the
CRAB technique for optimal quantum control that was originally
developed in quantum information science \cite{DCM} and used it
to determine, for realistic experimental parameters \cite{exp}, pulse
shapes that allow for the preparation of arbitrary coherent
superpositions with high fidelity. We are also able to reduce the
impact of the random orientation of FMO complexes in typical
samples and we are able to optimize the readout of the system
to maximize state sensitivity. Finally, we indicated that,
in a future experiment applying this approach, a comparison
of our theoretical calculations with experimental data, would
allow us to extract information about the largely unknown details
of the system-environment interaction, thus complementing
recently proposed methods based on tomography \cite{alan2011}. These
methods may support the experimental confirmation that recent
models concerning the interplay of transport processes and
environmental noise~\cite{Aspuru08,PlenioHuelga08,ccdhp09,castro08,ccdhp10,patrick,cdchp10,chp10} grasp the main features of the system
dynamics.

In future work, based on the techniques presented here, we
are planning to consider more general non-Markovian models~\cite{prior2010,chin2010,aki2009,thorwart2009},
other bio-molecular complexes and we will explore the
importance of multiple excitations in the system. We will also
explore the use of CRAB as a tool for closed-loop control in
this context. These tools will then be applied to provide optimized
experimental set-ups that explore questions concerning
the relevance of quantum coherence and decoherence in the
dynamics of bio-molecular systems. This includes the quantitative
probing of the functional relevance of quantum coherence
and entanglement, the exploration of which would shed
further light on the question whether entanglement is a necessary
ingredient for excitation energy transport.

\

\begin{acknowledgments}
We acknowledge discussions with Jon Marangos concerning experimental parameters. This work was supported by the EU Integrating projects Q-ESSENCE and AQUTE, the EU STREP project CORNER, SFB/TRR21, and the Alexander von Humboldt Foundation. F.C. was supported also by a Marie-Curie Intra-European Fellowship within the 7th European Community Framework Programme. We acknowledge the bwGRiD for computational resources.
\end{acknowledgments}
\end{document}